\documentclass[12pt]{article}

\usepackage[hmargin=0.75in,vmargin=0.75in]{geometry}
\usepackage{amsfonts}
\usepackage{amsmath}
\usepackage{amsthm}
\usepackage{algorithm}
\usepackage{algpseudocode}
\usepackage[english]{babel}
\usepackage{bm}
\usepackage[]{color}
\usepackage[mathscr]{eucal}
\usepackage{fancyhdr}
\usepackage{float}
\usepackage[T1]{fontenc}
\usepackage[]{graphicx}
\usepackage{hyperref}
\usepackage[utf8]{inputenc}
\usepackage{mathtools}
\usepackage{neuralnetwork}
\usepackage{setspace}
\usepackage{soul}
\usepackage{subfigure}
\usepackage{tikz}
\usepackage{titling}
\usepackage[round,authoryear]{natbib}
\usepackage{adjustbox}
\usepackage{multirow}
\usepackage{colortbl}
\usepackage{xcolor}
\usepackage{caption}
\usepackage{lineno}

\newlength{\subfigspacing}
\definecolor{lightblue}{rgb}{0.8,0.8,1}
\definecolor{lightgreen}{rgb}{0.8,1,0.8}

\newcommand{\blind}{0} 

\def\spacingset#1{\renewcommand{\baselinestretch}%
{#1}\small\normalsize} \spacingset{1}

\renewcommand{\hat}[1]{\widehat{#1}}
\renewcommand{\bar}[1]{\overline{#1}}

\newtheorem{theorem}{Theorem}
\DeclarePairedDelimiter{\norm}{\lVert}{\rVert}

\begin{document}
\bibliographystyle{chicago}


\if0\blind
{
  \title{\bf Dimension-reduced Reconstruction Map Learning for Parameter Estimation in Likelihood-Free Inference Problems}
  \author{Rui Zhang\\ 
    Department of Statistics, The Ohio State University\\
    and \\
    Oksana Chkrebtii \\
    Department of Statistics, The Ohio State University\\
    and \\
    Dongbin Xiu \\
    Department of Mathematics, The Ohio State University
    }
  \maketitle
} \fi

\if1\blind
{
  \bigskip
  \bigskip
  \bigskip
  \begin{center}
    {\LARGE\bf Title}
\end{center}
  \medskip
} \fi

\bigskip
\begin{abstract}
Many application areas rely on models that can be readily simulated but lack a closed-form likelihood, or an accurate approximation under arbitrary parameter values. Existing parameter estimation approaches in this setting are generally approximate. Recent work on using neural network models to reconstruct the mapping from the data space to the parameters from a set of synthetic parameter-data pairs suffers from the curse of dimensionality, resulting in inaccurate estimation as the data size grows. We propose a dimension-reduced approach to likelihood-free estimation which combines the ideas of reconstruction map estimation with dimension-reduction approaches based on subject-specific knowledge. We examine the properties of reconstruction map estimation with and without dimension reduction and explore the trade-off between approximation error due to information loss from reducing the data dimension and approximation error. Numerical examples show that the proposed approach compares favorably with reconstruction map estimation, approximate Bayesian computation, and synthetic likelihood estimation.  
\end{abstract}

\noindent%
\vfill

\newpage
\spacingset{1.75} 

\section{Introduction}
\label{sec:intro}

Statistical inference on dynamical systems, their latent parameters, and states, is critical for model assessment, interpretation, and prediction. However, the absence of a closed-form likelihoood makes likelihood-based or Bayesian inference infeasible. Models without a closed-form representation of the data-generating mechanism arise naturally in many modern application areas. Generative models can reflect the random stochastic nature of processes such as human interaction \citep[e.g.,][]{Kypraios2017, Chkrebtii2022,Chernozhukov2007}, the interaction between biological agents \citep[e.g.,][]{Kendall1999,Ashyraliyev2009,Auchincloss2008,Gilbert2008}, and reactions of chemical species \citep[e.g.,][]{Singer2006}. Although simulation from such models is possible and often computationally efficient, the likelihood often cannot be written down. 
Complex data types often lead to likelihoods with a combinatorially large number of components, such as interacting atomic spins on lattices \citep[e.g.,][]{Ghosal2020,Atchadé2013} and social networks \citep[e.g.,][]{Stivala2020}, or an intractable normalizing constant, such as probability models defined on a manifold \citep[e.g.,][]{Matuk2021}, Gaussian random fields \citep[e.g.,][]{Varin2011}, protein design \citep[e.g.,][]{Kleinman2006} and images \citep[e.g.,][]{Ibáñez2003}. In some cases, the likelihood is intractable due to latent variables in the data-generating model, such as for state space models \citep[e.g.,][]{Durbin2012}, hidden Markov models \citep[e.g.,][]{Yildirim2015}, mixed and random effects models \citep[e.g.,][]{Varin2011}, where the likelihood is a high-dimensional integral or summation over all latent variable values.

There are several popular approaches for statistical inference on models with intractable likelihood. However, for general models, such techniques typically require some degree of approximation. Composite likelihood methods \citep{Lindsay1988,Besag1975} approximate the likelihood by the product of lower-dimensional marginal or conditional densities. The construction of the composite likelihood components may be difficult, especially in complex models with many unknowns and, in general, the approximation may introduce non-negligible estimation bias \citep{Zhou2009, Friel2004}. 
In special cases where an unbiased estimator of the likelihood is available, the pseudo-marginal approach of \citet{Andrieu2009} enables exact Bayesian inference by replacing the likelihood evaluation within the Metropolis-Hastings algorithm. 
In addition to the method's lack of generality, the resulting MCMC sampler is often computationally inefficient, such as when the unbiased estimator itself requires a sampling algorithm \citep[e.g.,][]{Fallaize2016}. 
In contrast, the class of simulation-based estimation methods does not require point-wise evaluation of the likelihood if model output can be generated relatively quickly. 
A popular simulation-based approach is approximate Bayesian computation (ABC) \citep{Tavare1997, Pritchard1999, Beaumont2002}. ABC refers to the class of sampling techniques that target an approximate posterior distribution \citep{Fearnhead2012}, termed the ABC posterior, obtained by replacing the likelihood with a kernel density approximation based on the discrepancy between summarized synthetic and observed data. Since sufficient summary statistics are not typically available for likelihood-free problems, the choice of summary involves the trade-off between approximation and Monte Carlo errors. Using fewer summaries increases the approximation error between the ABC and the true posteriors due to information loss, while decreasing Monte Carlo approximation error as likelihood estimation becomes more efficient. 
Another popular simulation-based approach is synthetic likelihood estimation \citep{Wood2010}, which replaces the likelihood by a multivariate normal density with mean and covariance estimated from synthetic data. Although this approach scales well with data dimension, model misspecification can lead to estimation bias. 

 We propose a new simulation-based approach that utilizes neural networks (NN) to learn the mapping between observed data and model parameters from a large number of parameter-output pairs by exploiting dimension reduction. The advantage of using NNs is that they are universal function approximators \citep{Hornik1989} and have the flexibility to capture nonlinear relationships between variables.  
 NNs have been used for parameter estimation as a means of speeding up optimization, which is fundamentally different than our proposal. For instance, \citet{Morshed1998} use NNs as surrogate models trained on synthetic data, then perform optimization using a genetic algorithm. \citet{Matsubara2006} use a radial basis function network to learn the relationship between parameters and fitness value, then employ an optimization algorithm to find the setting that produces maximum fitness value. NNs can also be trained with synthetic data to learn conditional density estimators based on mixtures of Gaussian, normalizing flows or autoregressive flows as a surrogate model for the simulator, and the NN can either learn the posterior distribution \citep{Lueckmann2017, Papamakarios2016} or the likelihood \citep{Papamakarios2019, Alsing2019}. But this approach requires fitting a sequence of NN models to a possibly prohibitively large number of model evaluations, and is more computationally expensive than our proposal. 
 
 Our approach shares a foundation with recent literature on what we shall call reconstruction map (RM) estimation. \citet{Rudi2021} consider parameter estimation for the FitzHugh–Nagumo model by learning the mapping from the sample space to parameter space using a deep NN trained on a large number of synthetic datasets generated from the model. Similarly, \citet{Gerber2021} demonstrate that neural networks can be used to learn a mapping from  moderate-sized spatial fields to Gaussian process covariance parameters, offering a fast alternative to computationally demanding maximum likelihood estimation. \citet{Lenzi2021} further expand its application to intractable models with an example of parameter estimation for max-stable processes, showing its flexibility in handling highly non-Gaussian and spatially dependent data. \citet{SainsburyDale2024} complement this line of work by framing the reconstruction map approach as a direct approximation of the Bayes estimator—referred to as neural Bayes estimators, and by extending it to settings with independent replicates using permutation-invariant neural networks. The effectiveness of this approach is further demonstrated through additional simulation studies, including applications to a spatial conditional extremes model. Crucially, their approach suffers from the curse of dimensionality, i.e., its estimation performance degrades quickly as the data size grows. In an application for econometric models, \citet{Creel2017} proposes to use informative statistics as input to train a neural network, the output of which can be used directly as an estimator, or as an input to subsequent classical or Bayesian inference estimation. In a similar spirit, \citet{Rai2024} propose using a set of extreme quantiles as approximate sufficient statistics as input to a neural network for parameter estimation in the generalized extreme value (GEV) distribution. More literature on neural network-based methods for inference can be found in the recent review by \citet{ZammitMangion2025}, which provides a comprehensive overview of methodological developments in this area. 
 Our work builds on Creel's approach by establishing a systematic simulation-based Reconstruction Map-dimension Reduction (RM-DR) estimation method which overcomes the fundamental problem of degraded estimation performance with data dimension. We show that under certain assumptions, the resulting estimator is asymptotically equivalent to a Bayes estimator. Through multiple numerical experiments, we show that dimension reduction is essential for estimation from large datasets. We further propose a combined parameter estimation approach that utilizes the RM-DR as a starting point in a local optimization algorithm when the likelihood is available, providing an alternative to computationally costly global optimization methods. 

The rest of the paper is organized as follows. Section 2 introduces the inference problem and background required for constructing estimates of the reconstruction map. Section 3 establishes our approach, discusses its properties and describes criteria for assessing estimation accuracy. Section 4 discusses the results involving parameter estimation in four numerical experiments, three of which have an intractable likelihood, comparing the proposed approach with existing alternatives. In Section 5 we make conclusions and propose open questions for future work.


\section{Background}\label{background}
We begin by reviewing neural network models, which will later be used to construct estimators. We then describe the framework of reconstruction map estimation and point out drawbacks to its use when sample sizes are not low-dimensional.

\subsection{Neural Network Models}
We now review the basics of \textit{neural network} (NN) modelling and fitting. Broadly speaking, a neural network is a computational model made up of interconnected artificial nodes or \textit{neurons} in a layered structure that is intended to mimic the way the human brain works. A NN takes a given number of \textit{input variables}, processes them through one or more \textit{hidden layers}, and provides output in the \textit{output layer}. For example, in the context of a regression problem, the inputs to the NN are the training covariates or temporal indexes, and the output is the regression function. Similarly, in an image classification problem, the input is a training image and the output is a class label. 

\begin{figure}
	\begin{center}
		\subfigure{
                \scalebox{0.6}{
			\begin{tikzpicture}[>=latex]
                \path
                (0,0)     node[circle,draw,scale=1.2,inner sep=2pt] (S) {$x^{out}$}
                +(90:2.5) node[circle,draw,inner sep=2.5pt] (b) {}
                          node[above=1mm] {$b$}
                +(-3.5,2)  node[circle,draw]  (x1) {$\; x_1 \;\;$}
                +(-3.5,0.6)    node[circle,draw]  (x2) {$\; x_2\;\;$}
                +(-3.5,-0.2)    node {$\cdot$}
                +(-3.5,-0.6)    node {$\cdot$}
                +(-3.5,-1)    node {$\cdot$}
                +(-3.5,-2) node[circle,draw]  (x3) {$x_{n^{in}}$}
                (5,0)    node[scale=1.2] (g) {$x^{out}$};
                \draw[->] (S)--(g);
                \draw[->] (b)--(S);
                \draw[->] (x1)--(S) node[pos=.4,above]{$w_1$};
                \draw[->] (x2)--(S) node[pos=.4,above]{$w_2$};
                \draw[->] (x3)--(S) node[pos=.4,above]{$w_{n^{in}}$};
                \end{tikzpicture}
                }
                }
                \hspace{0.1in}
		\subfigure{%
                \scalebox{0.6}{
			\tikzstyle{mynode}=[thick,draw,circle,minimum size=22]
                \begin{tikzpicture}[x=2.2cm,y=1.4cm]
                  \foreach \N [count=\lay,remember={\N as \Nprev (initially 0);}]
                               in {3,4,4,4,2}{ 
                    \foreach \i [evaluate={\y=\N/2-\i; \x=1.5*\lay; \prev=int(\lay-1);}]
                                 in {1,...,\N}{ 
                      \node[mynode] (N\lay-\i) at (\x,\y) {};
                      \ifnum\Nprev>0 
                        \foreach \j in {1,...,\Nprev}{ 
                          \draw[->] (N\prev-\j) -- (N\lay-\i);
                        }
                      \fi
                    }
                  }
                  \node[above=5,align=center] at (N1-1.90) {input layer};
                  \node[above=2,align=center] at (N2-1.90) {$1^{\text{st}}$ hidden layer};
                  \node[above=2,align=center] at (N3-1.90) {$2^{\text{nd}}$ hidden layer};
                  \node[above=2,align=center] at (N4-1.90) {$3^{\text{rd}}$ hidden layer};
                  \node[above=8,align=center] at (N5-1.90) {output layer};
                \end{tikzpicture}
                }
		}
        \vspace{-0.15in}
	\end{center}
	\caption{Left: a neuron and its components; $x_1, \ldots, x_{n^{in}}$ are input nodes/variables with weights $w_1,\ldots, w_{n^{in}}$; $b$ is the bias parameter; $f$ is the activation function; and $x^{out}$ is the output. Right: a 4-layer neural network with 3 input nodes, 4 neurons per hidden layer, and 2 output neurons.}
	\label{fig:neuron_structure}
\end{figure}
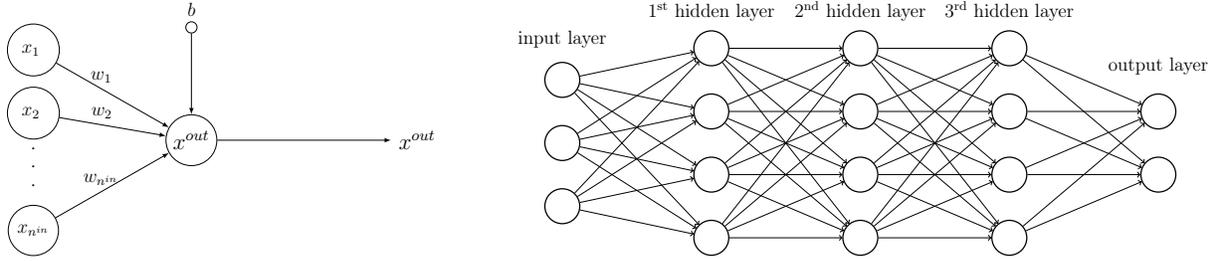

The key building block of the NN is the neuron as shown in the left panel of Fig. \ref{fig:neuron_structure}, which consists of $n^{in}$ input nodes/variables with the $i$th variable denoted as $x_i$ and associated \textit{weight parameter} $w_i$, a \textit{bias parameter} $b$, an \textit{activation function} $f:\mathbb{R}\rightarrow \mathbb{R}$, and the neuron output $x^{out}$. Let $x^{in}$ and $w$ be vector representations of input variables and weights, respectively. The output $x^{out}$ is calculated as evaluating the activation function at the weighted sum of input variables,
\begin{equation}
        x^{out}=f(w^\top x^{in}+b).
\end{equation}
%
%
A \textit{fully-connected layer} is formed by combining multiple neurons together. Let $x^{in}$ still be the vector of input variables, and $x^{out}$ be the vector of values for $n^{out}$ output neurons. Let $W:=(w_1,\ldots,w_{n^{out}})$ be the weight matrix, and $b:=(b_1,\ldots,b_{n^{out}})$ be the bias matrix which are formed by stacking weight vector and bias parameter, respectively, for each neuron horizontally. Define $A:=(W^\top, b^\top)$ as the matrix of all parameters in the layer and an affine linear function $h_{A}(x):=A\begin{pmatrix}x \\ 1\end{pmatrix}$. Then the output for this single layer is
\begin{equation}
        x^{out}=f\circ h_{A}(x^{in}),
\end{equation}
where the activation function $f$ is applied component-wise. 

More complex models can be constructed using multi-layer NNs, where there is a sequence of hidden layers between input and output layer as shown in the right panel of Fig. \ref{fig:neuron_structure}. For an L-layer NN,  denote $A_{l}$ as the matrix of all parameters in the $l$th layer, and $f_l$ as the activation function in $l$th layer, for $l=1,\ldots,L$. Let $\omega:=(A_{1},\ldots,A_{L})$ be parameters of the NN represented as a sequence of matrices, and $\mathbf{N}(\cdot,\omega):\mathbb{R}^{n^{in}} \rightarrow \mathbb{R}^{n^{out}}$ be the vector-valued function representation of the NN, 
\begin{equation}
        \mathbf{N}(\cdot,\omega)=f_L\circ h_{A_L}\circ\cdots \circ f_1\circ h_{A_1},  \label{NN operator}
\end{equation}
which is a composition of a series of alternating linear functions and activation functions with output
\begin{equation}
        x^{out}=\mathbf{N}(x^{in},\omega).
\end{equation}
The activation function is the key component that produces non-linearity of the NN. It must be monotonic and differentiable, as well as computationally inexpensive to evaluate along with its derivative. Commonly used activation functions include sigmoid, tanh, softplus, ReLU functions. The activation functions used in the hidden layers should be nonlinear, and ReLU function defined as $f(x)=\text{max}\{0,x\}$ has become a popular default choice because of its computational efficiency and representational sparsity. Its use leads to fast convergence for fitting NNs and mitigation of the vanishing gradient problem \citep{Glorot2011}. In the output layer, it is appropriate to set the activation function to be identity, as it can avoid undesirable constraints. The \textit{depth} of a NN is equal to the number of hidden layers $L-1$, and a NN is called deep if it has at least 2 hidden layers. When depth increases, the \textit{capacity} (model complexity) of a NN increases. Taken together, the above modeling choices comprise an \textit{architecture} or structure of the NN. The choice of architecture is largely problem-specific and involves a series of trade-offs between complexity and computational speed.

\subsection{Reconstruction Map (RM) Estimation}\label{RM estimator intro}

Let the observed data $y\in \mathbb{R}^m$ be a sample from a generative model that depends on unknown parameters $\theta\in \Theta\subseteq \mathbb{R}^d$, with likelihood function denoted as $p(y\mid \theta)$, but not necessarily known. We require that synthetic data $y$ can be readily simulated from $p(y\mid \theta)$ for arbitrary values of $\theta\in \Theta$, even if the likelihood $p(y\mid \theta)$ is computationally intractable. 
An estimator defines a mapping $\hat{\theta}:\mathbb{R}^m\rightarrow \Theta$ from the sample space to the parameter space. The popular likelihood-free estimation techniques ABC and SLE are reviewed in the supplement.

In this section, we describe a simulation-based method which we will call \textit{reconstruction map} (RM) estimation first proposed by \citet{Rudi2021} to estimate parameters defining an ODE from time series data. RM employs supervised learning to recover the mapping from the sample space to the parameter space based on a large number of synthetic datasets (or \textit{synthetic training data}) simulated from the data-generating model. Specifically, the mapping is modeled by a neural network, using the data $y$ as inputs and model parameters $\theta$ as outputs. The associated synthetic data defines the loss function used to train the NN model. Next, we present the details of RM estimation.

We denote by $d(\theta)$ a \textit{design density} function over $\theta$, which is used to generate $N$ synthetic training data-parameter pairs $(\theta_{n},y_{n})_{n=1}^{N}$, where  $(\theta_{n},y_{n})\stackrel{\text{ind}}{\sim}d(\theta)p(y\mid \theta)$ and $y_n \in \mathbb{R}^m$. For a given NN architecture, we denote the vector-valued NN function as $\mathbf{N}(\cdot,\omega):\mathbb{R}^m \rightarrow \mathbb{R}^{d}$, which is defined as in (\ref{NN operator}) and has parameters $\omega$. The loss function is denoted as $l(\cdot,\cdot)$, with $l(\theta,\hat{\theta})$ representing the loss associated with an estimate of  $\hat{\theta}$. 
The estimation performance for the NN is assessed via the training loss $\frac{1}{N} \sum_{n=1}^{N}l(\theta_{n},\mathbf{N}(y_{n},\omega))$. 
RM estimation trains the NN until a maximum number of epochs is reached, and taking the estimator to be the NN function with parameters that minimize the training loss over $\omega\in \Omega$. That is, the RM estimator is
\begin{equation}
\hat{\theta}_{RM}(y)=\mathbf{N}(y,\omega^{*}), \; \text{ where }\omega^{*}\in \underset{\omega\in\Omega}{\operatorname{argmin}} \frac{1}{N} \sum_{n=1}^{N}l(\theta_{n},\mathbf{N}(y_{n},\omega)).  \label{RM est}
\end{equation}
The computational implementation of the RM estimator is presented in Algorithm \ref{RM algo}.

\begin{algorithm}
\caption{Algorithm for RM estimation}\label{RM algo}
\begin{algorithmic}[1]
    \Require $\text{ design density }d(\cdot), \text{ data-generating process with density }p(\cdot\mid \cdot),\text{ NN model } \mathbf{N}(\cdot,\cdot), $
    \Statex $\text{ loss function }l(\cdot,\cdot),\text{ integer } N>0$ 
\Ensure $\hat{\theta}_{RM}(\cdot)$
\For{$n=1$ to $N$}
\State sample $\theta_{n}\sim d(\cdot)$
\State sample $y_{n}\mid \theta_{n} \sim p(\cdot\mid \theta_{n})$
\EndFor
\State use numerical optimization to solve $\omega^{*}\in\underset{\omega\in\Omega}{\operatorname{argmin}} \frac{1}{N} \sum_{n=1}^{N}l(\theta_{n},\mathbf{N}(y_{n},\omega))$
\State set $\hat{\theta}_{RM}(\cdot)=\mathbf{N}(\cdot,\omega^{*})$
\end{algorithmic}
\end{algorithm}

While likelihood-based estimation methods such as MLE or Bayes estimation require evaluating the likelihood at arbitrary locations, the RM estimation method is substantially different in that it is likelihood-free and only requires being able to simulate data from the generating model. 
Another desirable feature of RM estimation is that estimation from new data under the same generating model only requires evaluating the pre-learned reconstruction map. In contrast, popular likelihood-free methods including ABC and SLE require repeating the entire algorithm as new data arrives. 

As a simulation-based approach, a notable difference between RM estimation and ABC or SLE is that RM estimation does not require data summarization, as it provides the full data as input to the neural network. RM estimation essentially attempts to learn the key parts of the data and summarize it into features automatically through neurons in hidden layers by training the neural network with synthetic training data. But one important drawback of RM estimation is that its performance quickly degrades as the dimension of the data grows, i.e., the input space of the reconstruction map becomes large. This issue, which will be explained in detail in the following section, makes the approach originally proposed by \citet{Rudi2021} infeasible in all but relatively small data problems.  In the following section, we will introduce the new \textit{dimension-reduced reconstruction map} (RM-DR) estimation technique that resolves this problem by incorporating dimension reduction of the input space, establish its connection with Bayes estimation, and analyze different sources of estimation error.


\section{Methodology}\label{method}
This section introduces our simulation-based \textit{dimension-reduced reconstruction map} (RM-DR) estimator and discusses its properties. Uncertainty quantification is further discussed in the supplementary materials. In addition to RM-DR estimation, we also propose the RM-DRLO method that facilitates likelihood-based inference when the likelihood is available but potentially expensive to evaluate. Details are provided in the supplement. Finally, we discuss criteria to evaluate and compare estimators in the likelihood-free setting. 

\subsection{Dimension-reduced Reconstruction Map (RM-DR) Estimator}\label{RM-DR estimator intro}
Section \ref{RM estimator intro} discussed how a supervised learning technique may be used to learn the reconstruction map from the data to the parameter space in order to construct estimators when the likelihood is not available. RM estimation can be viewed as learning the manifold that describes the relationship between the observed data and the parameters defining the generative model. However, in contrast to most standard statistical methods for which estimation performance grows with the observed data dimension, RM approximation degrades as the dataset grows due to the increasing dimension of the manifold's input domain. Counteracting this effect requires potentially infeasible increases in the training data size. The proposed RM-DR estimation approach resolves this problem by projecting both the observed and synthetic data into a low-dimensional space before learning the reconstruction map, resulting in a lower-dimensional manifold that is easier to learn. The effect of dimension reduction can be understood as a trade-off between two types of error for RM-DR estimation: information loss due to summarization of the full data, and approximation error associated with learning the dimension-reduced manifold from synthetic data. As the dimension of the summary statistic decreases, the information loss increases while the approximation error decreases. 
While RM estimation does not suffer from such compression error, even relatively small datasets will result in large Monte Carlo approximation error. Thus, reducing the input dimension of the reconstruction map enables the use of RM-DR with high-dimensional data. An additional benefit of dimension reduction, such as smoothing a time series, could denoise the observed data, revealing key features that are informative about the model parameters. Finally, RM-DR demonstrates better performance in simulation studies relative to RM and adheres to the principle of parsimony. We will further discuss the role of dimension reduction in Section \ref{property}.

An important practical advantage of the proposed RM-DR estimation approach is its computational efficiency relative to likelihood-free methods such as ABC and SLE. While RM-DR requires an upfront computational cost during training (involving simulation of synthetic datasets and neural network optimization), this cost is amortized across future inferences. Once the reconstruction map is learned, estimation from a new dataset under the same generative model becomes highly efficient, requiring only a forward evaluation of the pre-trained neural network and dimension reduction. Importantly, no further simulations from the model are required during inference, making the per-dataset inference cost effectively \(O(1)\) with respect to the number of model simulations.
In contrast, ABC requires simulating synthetic datasets until a sufficient number of accepted samples are obtained that match the observed data within a specified tolerance, leading to a computational complexity of \(O(N_{\text{post}})\), where \(N_{\text{post}}\) is the desired number of posterior samples. For SLE, each likelihood evaluation requires simulating \(N_s\) synthetic datasets to estimate the synthetic likelihood, and the optimization typically requires \(N_{\text{iter}}\) iterations, resulting in a computational complexity of \(O(N_s \cdot N_{\text{iter}})\) per dataset. Consequently, RM-DR achieves estimation at a substantially lower computational cost than ABC or SLE when inference is performed on multiple or many datasets under the same model, making it well-suited for large-scale likelihood-free inference tasks where repeated parameter estimation is required.



Suppose that $s=S(y)$ denotes a summary $S:\mathbb{R}^{m}\rightarrow \mathbb{R}^{K}$ that reduces the data dimension from $m$ to $K<m$. As in RM estimation, RM-DR defines a design distribution on the model parameters with density $d(\cdot)$. We will discuss interpretation and optimal choices of $d(\cdot)$ in Section \ref{property}.  The $N$ synthetic training data-parameter pairs are $(\theta_{n},s_{n})_{n=1}^{N}$, where $(\theta_{n},y_{n})\sim p(y\mid \theta)$, and $s_{n}:=S(y_{n})$,  for $n=1,\ldots, N$. 
Similarly to Section \ref{RM estimator intro}, the vector-valued NN function is denoted as $\mathbf{N}(\cdot,\omega):\mathbb{R}^K \rightarrow \mathbb{R}^{d}$, where $\omega$ are the parameters defining the NN. And the training loss is $\frac{1}{N} \sum_{n=1}^{N} l(\theta_n,\mathbf{N}(s_n,\omega))$. 
A NN is trained until a user-specified maximum number of epochs is reached, determined based on pilot experiments, domain knowledge, or heuristic guidelines, or until the training loss does not substantially decrease across a fixed number of epochs. The RM-DR estimator is given by the NN with parameters minimizing the training loss over $\omega\in \Omega$,

\begin{equation}
\hat{\theta}_{RMDR}(s)=\mathbf{N}(s,\omega^{*}), \; \text{ where } 
 \omega^{*}\in \, \underset{\omega\in\Omega}{\operatorname{argmin}} \, \frac{1}{N} \sum_{n=1}^{N}l(\theta_n,\mathbf{N}(s_n,\omega)). \label{RM-DR est}
\end{equation}
The implementation of the RM-DR estimation procedure is presented in Algorithm \ref{RM-DR algo}. 

\begin{algorithm}
\caption{Algorithm for RM-DR estimation}\label{RM-DR algo}
\begin{algorithmic}[1]
\Require $\text{ design density }d(\cdot), \text{ data-generating process with density }p(\cdot\mid \cdot),\text{ NN model } \mathbf{N}(\cdot,\cdot), $
    \Statex $\text{ summary function } S(\cdot), \text{ loss function }l(\cdot,\cdot),\text{ integer } N>0$ 
\Ensure $\hat{\theta}_{RMDR}(\cdot)$
\For{$n=1$ to $N$}
\State sample $\theta_{n}\sim d(\cdot)$
\State sample $y_{n}\mid \theta_{n} \sim p(\cdot\mid \theta_{n})$
\State calculate $s_{n}=S(y_{n})$
\EndFor
\State use numerical optimization to solve $\omega^{*}\in\underset{\omega\in\Omega}{\operatorname{argmin}} \frac{1}{N} \sum_{n=1}^{N}l(\theta_n,\mathbf{N}(s_n,\omega))$
\State set $\hat{\theta}_{RMDR}(\cdot)=\mathbf{N}(\cdot,\omega^{*})$
\end{algorithmic}
\end{algorithm}

A practical issue that arises when fitting NN models to data is that of over-fitting. In order to avoid this issue for both RM and RM-DR methods, we suggest first generating \textit{synthetic validation data} in the same manner as the remaining training data. This validation data is used to determine the optimization algorithm's stopping time (maximum number of epochs), by minimizing the validation loss rather than the training loss. We employ this approach in all of our numerical experiments. 

\subsection{Connection with Bayes Estimation}\label{property}
In this section, we establish a connection between the RM-DR estimator and the Bayes estimator. As discussed earlier, the dimension-reduced data, $s\in \mathcal{S}\subseteq \mathbb{R}^{K}$, is used as input in the reconstruction map to estimate $\theta\in \Theta \subseteq \mathbb{R}^d$, where $\mathcal{S}$ denotes the support of $s$. For a given NN architecture, denote the set of vector-valued NN functions as $\mathcal{A}=\{\mathbf{N}(\cdot,\omega)\mid \omega\in \Omega\}$, where $\mathbf{N}(\cdot,\omega):\mathbb{R}^{K} \rightarrow \mathbb{R}^{d}$. To define a Bayes estimator, we require a prior distribution $\pi(\cdot)$ on the parameters. For an estimator $g:\mathbb{R}^{K} \rightarrow \mathbb{R}^{d}$, Bayes risk is defined as 
\begin{equation}
   r_{s}(\pi,g):=\mathbb{E}_{(s,\theta)\sim p_{\pi}(s,\theta)}\bigl[ l(\theta,g(s))\bigr],  \label{Bayes risk def}
\end{equation}
where the expectation is taken over the joint distribution with density $p_{\pi}(s,\theta)=p(s\mid \theta)\pi(\theta)$. An estimator $\hat{\theta}_{B}$ is a Bayes estimator if it minimizes the Bayes risk among all estimators: 
\begin{equation}
    \hat{\theta}_{B}\in\underset{g}{\operatorname{argmin}}\;  r_{s}(\pi,g).  \label{Bayes estimator}
\end{equation}

Let $Q_n(\omega) = \frac{1}{n} \sum_{i=1}^{n} l(\theta_i, \mathbf{N}(s_i, \omega))$ be the training loss function, where \( (\theta_i, s_i)_{i=1}^{n} \) are i.i.d. synthetic training data-parameter pairs from the joint distribution with density \( p_d(s, \theta) = p(s \mid \theta) d(\theta) \). Since \( (\theta_i, s_i) \) are random, \( Q_n(\omega) \) is a random function of \( \omega \), with randomness induced by the synthetic training data. Define the expected training loss function as $Q_0(\omega)=\mathbb{E}_{(s,\theta)\sim p_{d}(s,\theta)}\bigl[Q_n(\omega)\bigr] = \mathbb{E}_{(s,\theta)\sim p_{d}(s,\theta)}\bigl[l(\theta,\mathbf{N}(s, \omega))\bigr]$.  In this section, we define the RM-DR estimator as  
\[
\hat{\theta}_n(\cdot) = \mathbf{N}(\cdot, \hat{\omega}_n), \quad \text{where} \quad \hat{\omega}_n \in \underset{\omega \in \Omega}{\operatorname{argmin}} \; Q_n(\omega).
\]
The following theorems formalize the connection between RM-DR and Bayes estimation. Detailed proofs are provided in the supplement.

\begin{theorem}\label{consistency}
Assume that:
\begin{enumerate}
    \item The space $\Omega$ of parameters defining the neural network is compact;
    \item The neural network function $\mathbf{N}(s, \omega)$ is continuous in $\omega$ for any fixed $s\in \mathcal{S}$;
    \item The expected training loss function $Q_0(\omega)< \infty$ for any $\omega\in \Omega$ and has a set of minimizers $\Omega_0 = \underset{\omega\in\Omega}{\operatorname{argmin}} \, Q_0(\omega)$ such that for any \( \omega_a, \omega_b \in \Omega_0 \), 
    \(
        \mathbf{N}(\cdot, \omega_a) = \mathbf{N}(\cdot, \omega_b).
    \)
    That is, the induced NN function at the minimizers is unique, denoted as $\mathbf{N}_0(\cdot)$;
    \item The training loss function converges to the expected training loss function uniformly in probability: $\sup_{\omega \in \Omega} |Q_n(\omega) - Q_0(\omega)| \xrightarrow{p} 0$ as $n\to\infty$.
\end{enumerate}
Then, the RM-DR estimator $\hat{\theta}_n(\cdot)$ converges pointwise in probability to the function $\mathbf{N}_0(\cdot)$ as the number of synthetic training data-parameter pairs $n\to\infty$. That is, for each fixed $s\in \mathcal{S}$:
\[
    \hat{\theta}_n(s)\xrightarrow{p} \mathbf{N}_0(s), \text{ as }n \to \infty.
\]
Moreover, if we additionally assume that:
\begin{enumerate}
    \setcounter{enumi}{4}
    \item The support of summary statistics $\mathcal{S}$ is compact.
    \item The neural network function $\mathbf{N}(s, \omega)$ is jointly continuous in $(s, \omega) \in \mathcal{S} \times \Omega$,
\end{enumerate}
then the RM-DR estimator $\hat{\theta}_n(\cdot)$ converges uniformly in probability to the function $\mathbf{N}_0(\cdot)$ over $\mathcal{S}$ as $n\to\infty$. That is,
\[
    \sup_{s \in \mathcal{S}} |\hat{\theta}_n(s) - \mathbf{N}_0(s)| \xrightarrow{p} 0, \text{ as } n \to \infty.
\]
\end{theorem}

\begin{theorem}\label{convergence}
Suppose there exists a Bayes estimator $\hat{\theta}_{B}(\cdot)$ within $\mathcal{A} = \{\mathbf{N}(\cdot,\omega) : \omega \in \Omega\}$, and that the design density $d(\theta)$ and prior density $\pi(\theta)$ agree except on a set of Lebesgue measure zero. Further, suppose that Assumptions 1--4 in Theorem \ref{consistency} hold. Then the RM-DR estimator $\hat{\theta}_n(\cdot)$ converges pointwise in probability to the Bayes estimator $\hat{\theta}_{B}(\cdot)$ as the number of synthetic training data-parameter pairs $n\to\infty$. That is, for each fixed $s\in \mathcal{S}$:
\[
    \hat{\theta}_n(s)\xrightarrow{p} \hat{\theta}_{B}(s), \text{ as }n \to \infty.
\]
Moreover, if Assumptions 5-6 in Theorem \ref{consistency} are also satisfied, then the RM-DR estimator $\hat{\theta}_n(\cdot)$ converges uniformly in probability to the Bayes estimator $\hat{\theta}_{B}(\cdot)$ as $n \to \infty$. That is,
\[
    \sup_{s \in \mathcal{S}} |\hat{\theta}_n(s) - \hat{\theta}_{B}(s)| \xrightarrow{p} 0, \quad \text{as } n \to \infty.
\]
\end{theorem}



Clearly,
\[
Q_0(\omega) = \mathbb{E}_{(s,\theta)\sim p_{d}(s,\theta)}\bigl[l(\theta,\mathbf{N}(s, \omega))\bigr] = r_{s}(d,\mathbf{N}(\cdot, \omega))
\]
can be viewed as the Bayes risk for the estimator $\mathbf{N}(\cdot, \omega)$ with the design distribution $d(\cdot)$ playing the role of the prior. One implication of Theorem~\ref{consistency} is that under mild assumptions, the RM-DR estimator will converge in probability to an estimator that minimizes the Bayes risk over the class of functions specified by the neural network architecture. For a finite training sample size, the RM-DR estimator minimizes the empirical Bayes risk over the specified neural network class, with the design distribution serving as an analogue to the prior, thereby allowing the incorporation of prior knowledge about the parameter distribution.

A desirable property of Bayes estimators is that with respect to proper priors, they are virtually always admissible \citep{Berger1985}, meaning that there is no other estimator, as a function of $s$, that achieves a strictly smaller risk for every $\theta$. Theorem \ref{convergence} then shows that if the neural network class is sufficiently rich, under mild conditions, the RM-DR estimator becomes equivalent to the Bayes estimator as the training sample size grows large. This connection provides a theoretical justification for using RM-DR in practice, ensuring that as the neural network class becomes sufficiently expressive and the training sample size grows, RM-DR yields Bayes-optimal decisions under mild assumptions.



\subsection{Understanding Dimension Reduction}\label{dimensionreduction}

To better understand the effect of dimension reduction, for any estimator $\hat{\theta}: \mathbb{R}^{m} \rightarrow \mathbb{R}^{d}$, we denote the Bayes risk $r(\pi,\hat{\theta}):=\mathbb{E}_{(y,\theta)}\bigl[ l(\theta,\hat{\theta}(y))\bigr]$ and $ \hat{\theta}_{O}=\underset{\hat{\theta}}{\operatorname{argmin}} \;  r(\pi,\hat{\theta})$. And we denote $\hat{\theta}_{RMDR}^{S}(y)=\hat{\theta}_{RMDR}(S(y))$, $\hat{\theta}_{B}^{S}(y)=\hat{\theta}_{B}(S(y))$ and $g^{S}(y)=g(S(y))$. Based on (\ref{Bayes estimator}), equivalently we can write  $\hat{\theta}_{B}^{S}=\underset{g^{S}}{\operatorname{argmin}}\;  r(\pi,g^{S})$. For simplicity and without loss of generality, we assume $\hat{\theta}_{O}$ and $\hat{\theta}_{B}$ in (\ref{Bayes estimator}) are both unique, and $d(\theta)$ and $\pi(\theta)$ agree except on a measure of zero. Ideally, we want our estimator to be as close as possible to $\hat{\theta}_{O}$ that minimizes the Bayes risk among all estimators as functions of full data. And it is clear that the RM estimator converges to $\hat{\theta}_{O}$ if the number of synthetic training data-parameter pairs $N$ goes to infinity and the neural network is sufficiently expressive. However, in the finite sample case, the approximation error between $\hat{\theta}_{RM}$ and $\hat{\theta}_{O}$ is positive and may be large if the data dimension is high due to the inherent difficulty of estimating a function with a large input space. On the other hand, the RM-DR estimator will converge to $\hat{\theta}_{B}^{S}$, which minimizes the Bayes risk among all estimators based on $S(\cdot)$. Therefore in the finite sample case, the discrepancy between RM-DR estimator $\hat{\theta}_{RMDR}^{S}$ and the Bayes estimator $\hat{\theta}_{O}$ is composed of two types of error, one is the systematic error between $\hat{\theta}_{B}^{S}$ and $\hat{\theta}_{O}$, and the other is the  approximation error between $\hat{\theta}_{RMDR}^{S}$ and  $\hat{\theta}_{B}^{S}$. Compared with RM estimation, although RM-DR estimation has this systematic error, the approximation error is reduced due to a lower-dimensional input. So RM-DR is able to produce a lower aggregate error, and the overall effect becomes more pronounced as data dimension increases. In terms of degree of dimension reduction, there is a trade-off between the two types of error. Generally speaking, when the dimension  $K$ of the summary statistics decreases, the systematic error of the RM-DR estimator will increase due to loss of information, but the approximation error will decrease as it becomes easier to estimate a function with smaller input space. Theoretically, an estimator based on a low-dimensional sufficient statistic would be optimal since it would incur no systematic error and produce a lower approximation error that under the use of the full data. Unfortunately, finding a low-dimensional sufficient statistic in the likelihood-free setting is typically infeasible. Therefore, in practice, the choice of summaries is usually problem-specific and requires domain knowledge. In general, it is desirable for these summaries to reflect important features of data, and also depend on the unknown model parameters. For example, one may consider marginal distribution statistics such as sample moments, quantiles and order statistics. For time/spatially indexed data, we can consider descriptive features like the number of peaks or valleys, smoothness, shape of curves, frequency, amplitude, counts, etc.  Summaries of temporal or spatial dependence like auto-covariance may also be useful.

\subsection{Evaluating Estimation Performance}\label{fitcriteria}
To evaluate the performance of an estimator $\hat{\theta}: \mathbb{R}^{m} \rightarrow \mathbb{R}^{d}$, we first consider its risk, defined as 
\begin{equation}
    R(\theta,\hat{\theta}):=\mathbb{E}_{y\mid \theta}\bigl[l(\theta,\hat{\theta}(y)) \bigr],\label{risk} 
\end{equation}
which is its expected loss for a given $\theta$. To account for differences between relative performance of the estimator across the parameter space, we consider the Bayes risk as an aggregate measure of an estimator's expected error, defined as 
\begin{equation}
r(\rho,\hat{\theta}):=\mathbb{E}_{(y,\theta)}\bigl[ l(\theta,\hat{\theta}(y)) \bigr]=\mathbb{E}_{\theta\sim \rho}\bigl[ R(\theta,\hat{\theta})\bigr],\label{Bayes risk} 
\end{equation}
which averages the risk over a distribution $\rho(\cdot)$ on $\theta$. 
When no prior knowledge about $\theta$ is available, it would be appropriate to set $\rho(\cdot)$ as a uniform distribution over the parameter space. In practice, for RM and RM-DR, it would be reasonable to set the design distribution to be the same as $\rho$ to incorporate this uncertainty when designing our estimator. 

In practice, the risk and Bayes risk are not available in closed form, and a Monte Carlo (MC) approximation is used instead. We generate test data $\{(\theta_{q},\{y_{ql}\}_{l=1}^{L})_{q=1}^{Q}\}$, where  $\theta_{q}\stackrel{\text{i.i.d.}}{\sim}\rho(\cdot)$ for $q=1,\ldots, Q$, and $y_{ql}|\theta_{q} \stackrel{\text{i.i.d.}}{\sim} p(\cdot|\theta_{q})$ for $l=1,\ldots, L$, are replicates of the data generated under each $\theta_{q}$. For a given $\theta_q$, the MC approximation of the risk is
\begin{equation}
\frac{1}{L}\sum_{l=1}^{L}l(\theta_q,\hat{\theta}(y_{ql})),\label{MC risk} 
\end{equation}
and the MC approximation of the Bayes risk is
\begin{equation}
\frac{1}{QL}\sum_{q=1}^{Q}\sum_{l=1}^{L}l(\theta_q,\hat{\theta}(y_{ql})).\label{MC Bayes risk} 
\end{equation}
Under the commonly used squared error loss function, the risk is equal to the mean squared error (MSE), and the Bayes risk is referred to as integrated mean squared error (IMSE). Their MC approximations can be computed and decomposed as
\begin{equation}
\begin{split}
    \hat{\text{MSE}}(\theta_q,\hat{\theta})&=\frac{1}{L}\sum_{l=1}^{L}\norm{\theta_q-\hat{\theta}(y_{ql})}_2^2\\
    &=\norm{\theta_q-\bar{\hat{\theta}_q}}_2^2+\frac{1}{L}\sum_{l=1}^{L}\norm{\hat{\theta}(y_{ql})-\bar{\hat{\theta}_q}}_2^2, \label{MC MSE}
\end{split}
\end{equation}
and 
\begin{equation}
\begin{split}
    \hat{\text{IMSE}}(\rho,\hat{\theta})&=\frac{1}{QL}\sum_{q=1}^{Q}\sum_{l=1}^{L}\norm{\theta_q-\hat{\theta}(y_{ql})}_2^2\\
    &=\frac{1}{Q}\sum_{q=1}^{Q}\norm{\theta_q-\bar{\hat{\theta}_q}}_2^2+\frac{1}{QL}\sum_{q=1}^{Q}\sum_{l=1}^{L}\norm{\hat{\theta}(y_{ql})-\bar{\hat{\theta}_q}}_2^2,\label{MC IMSE}
\end{split}
\end{equation} 
where $\bar{\hat{\theta}_q}=\frac{1}{L}\sum_{l=1}^{L}\hat{\theta}(y_{ql})$. In (\ref{MC MSE}), $\norm{\theta_q-\bar{\hat{\theta}_q}}_2^2$ is the MC approximation of squared bias, and $\frac{1}{L}\sum_{l=1}^{L}\norm{\hat{\theta}(y_{ql})-\bar{\hat{\theta}_q}}_2^2$ is MC approximation of variance at $\theta_q$. In (\ref{MC IMSE}), $\frac{1}{Q}\sum_{q=1}^{Q}\norm{\theta_q-\bar{\hat{\theta}_q}}_2^2$ will be referred as MC approximation of integrated squared bias ($\hat{\text{IBIAS}}^2$), which represents average squared bias of the estimator, and $\frac{1}{QL}\sum_{q=1}^{Q}\sum_{l=1}^{L}\norm{\hat{\theta}(y_{ql})-\bar{\hat{\theta}_q}}_2^2$ will be referred to as MC approximation of integrated variance ($\hat{\text{IVAR}}$), which is the average variance of the estimator. In Section \ref{experiment}, we will use squared loss, and the criteria discussed above to evaluate an estimator's performance in numerical experiments.

\section{Numerical Experiments}\label{experiment}
We consider four simulated examples to evaluate the performance of the proposed estimation framework. The first three examples feature an intractable likelihood, while the final example is defined by a highly nonlinear and nonconvex likelihood surface, which poses computational challenges to likelihood-based methods. An accessible likelihood allows us to compare RM and RM-DR's performance with maximum likelihood estimation and to demonstrate the RM-DRLO approach in the supplement. The RM, RM-DR, and RM-DRLO approaches are based on a fully-connected neural network with 2 hidden layers, each having 32 neurons, and a ReLU activation function. The choice of this NN architecture involves trial and error through evaluating performance on synthetic validation data, and other more complex NN architectures including Convolutional Neural Networks (CNNs) and Long Short-Term Memory networks (LSTMs) produce comparable performance (which we do not show here for brevity). The design distribution for parameters is taken to be a uniform distribution over the parameter space. The training sample consists of 125,000 output-parameter pairs, of which $25\%$ is held out for validation. Evaluation is based on the fit criteria discussed in Section \ref{fitcriteria}, which are approximated based on $L = 100$ replications. ABC is implemented using adaptive tuning of the proposal covariance and a parallel tempering algorithm to enable efficient exploration of the ABC posterior \citep{Swendsen1986, Geyer1991}. A Gaussian kernel is used to measure the similarity between observed and synthetic data, with the bandwidth parameter chosen manually to be as small as possible while resulting in an acceptance rate within the target range. Convergence is assessed by monitoring traceplots and correlation plots. Finally, MLE and SLE estimators are obtained via numerical optimization using the dual annealing algorithm. 

\subsection{Ricker Model}
We first consider parameter estimation for the Ricker model, a discrete-time ecological model that describes the density-dependent dynamics of an animal population. The population density $N(t)$ is updated across a set of discrete time steps $t \in \mathbb{Z}^+$ via,
\begin{equation}
  N(t+1)=aN(t)e^{-N(t)+\epsilon(t)}, \label{Ricker model} 
\end{equation}
where $\epsilon(t)\stackrel{\text{ind}}{\sim}\mathcal{N}(0,\sigma^2)$ represents process noise within the dynamical system, and $a$ is an intrinsic growth rate parameter. Population size follows a Poisson model with mean $\delta N(t)$,
\begin{equation}
  y(t)\stackrel{\text{ind}}{\sim} \text{Poisson}(\delta N(t)), \label{Poisson observation} 
\end{equation}
where $\delta$ is an unknown scale parameter. The initial population is $N(0)=2$, and data $y=(y(1),\ldots,y(1,000))^\top$ is observed at $m = 1,000$ consecutive time steps. Setting $\eta=\log(a)$, the parameters of interest are $\theta=(\eta,\sigma,\delta)^\top \in (2,5)\times(0,0.3)\times(1,4)$. Supplement Fig. S1 shows four replications of $y$ under $\theta = (3, 0.2, 2)^\top$, illustrating the diversity of sample paths that are possible under the same parameter setting.  A likelihood calculation would require marginalization over $m$ unobserved population densities, and is thus effectively intractable. 


RM-DR, SLE, and ABC are implemented using the summary statistics suggested in \citet{Wood2010}. Let $\Delta(t)=y(t)-y(t-1)$ denote the differences between consecutive observations and let ${\Delta}_{(t)}$ be the $t$-th order statistics of $\Delta(1),\dots, \Delta(1000)$. Similarly, let $y_{(t)}$ be $t$-th order statistics of $y(1),\dots, y(1,000)$. 
The summary statistics are: the sample mean $\bar{y}=\frac{1}{1,000}\sum_{t=1}^{1000}y(t)$, sample autocovariance $\upsilon(h)=\frac{1}{1,000}\sum_{t=1}^{1,000-h}(y(t+h)-\bar{y})(y(t)-\bar{y})$ with lag $h$ from $0$ to $5$, number of zeros observed $\tau=\sum_{t=1}^{1,000} 1(y(t)=0)$, coefficients of the cubic regression of ordered differences $\Delta^{(t)}$ on the ordered observed values $y^{(t)}$, and coefficients of the autoregression of ${(y(t+1))}^{0.3}$ on ${y(t)}^{0.3}$ and ${y(t)}^{0.6}$. The rationale for these choices is as follows. The sample mean and autocovariance are typically useful summaries of time series data, while the frequency of zero observations can provide insights into the distribution of Poisson data. Coefficients of the cubic regression can summarize the marginal distribution of observations, and coefficients of the autoregression contain information about dynamic structure. 

\begin{figure}
	\begin{center}
		\subfigure{%
			
			\includegraphics[height=0.25\textwidth]{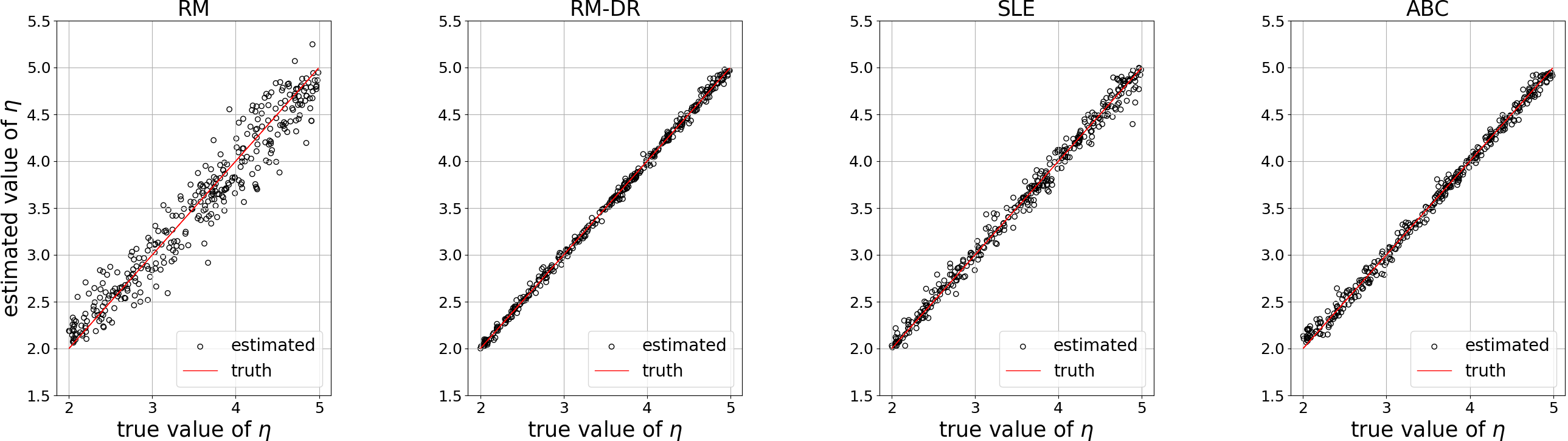}
		}\\
            \vspace{\subfigspacing}
		\subfigure{%
			
			\includegraphics[height=0.25\textwidth]{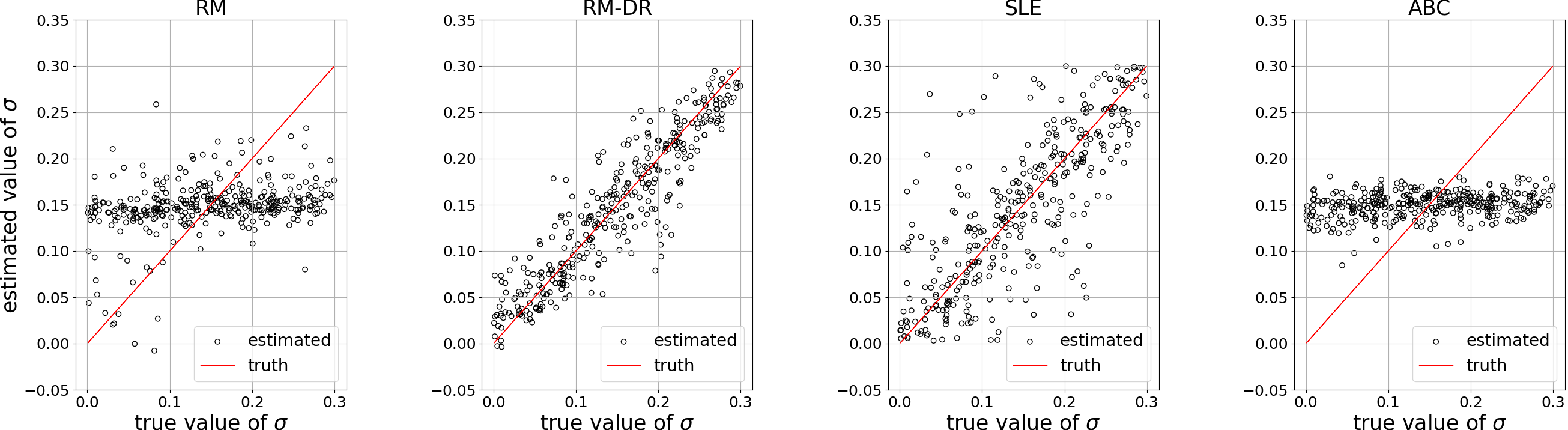}
		}\\
            \vspace{\subfigspacing}
		\subfigure{%
			
			\includegraphics[height=0.25\textwidth]{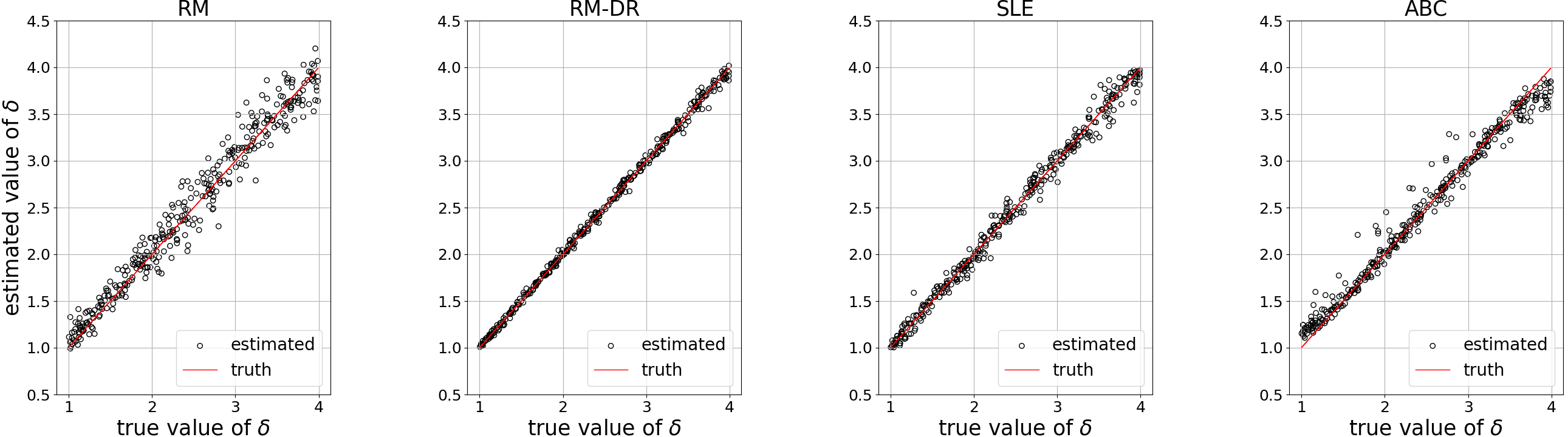}
		}
	\end{center}
        \vspace{-0.1in}
	\caption{Scatter plots of estimates versus simulation values of $\eta, \sigma$, and $\delta$  (rows), respectively, using RM, RM-DR, SLE, and ABC (columns), respectively, for the Ricker model example. The 45\textdegree\ line is shown in red for reference.}
	\label{Scatter Ricker}
\end{figure}

Fig. \ref{Scatter Ricker} shows scatter plots of estimates versus simulation values of $\eta, \sigma$ and $\delta$ (rows), respectively, under the four different estimation methods (columns). Out of the approaches considered, RM-DR estimates are the closest to true values for all components of $\theta$ (smallest spread around the 45\textdegree\ line, in red). RM achieves the worst performance among all methods considered. ABC shows comparable estimation accuracy to SLE, except for the estimation of the standard deviation $\sigma$. 
Supplement Fig. S2 
compares the performance of RM and RM-DR estimators using the evaluation criteria introduced in Section \ref{fitcriteria}. Each point on the 3-d plots corresponds to one of 1,000 different simulation parameter setting. The color corresponds to the magnitude of the log squared bias, variance, and MSE (rows), respectively, for RM (left column) and RM-DR (right column). RM-DR estimators achieve lower squared bias, variance and MSE across almost all parameter values compared to RM. Indeed, RM-DR  achieves substantially lower integrated squared bias, variance, and MSE. 
Due to the relatively expensive computation (approximately several hours for a single estimate), the performance of ABC and SLE are compared under three different $\theta$ settings in Table \ref{table Ricker}. RM-DR has the lowest MSE, variance, and squared bias in almost all cases, followed by ABC and SLE. RM has the worst performance over all metrics. 
In summary, this example illustrates that summarization of the data informing the RM-DR method greatly improves estimation performance relative to the RM approach, and performs favorably relative to ABC and SLE both in terms of accuracy and speed under the same choice of summaries.

\begin{table}
        \small
	\begin{center}
  \caption{MC approximation of squared bias, variance, and MSE for Ricker model example}
\label{table Ricker}	
		\begin{tabular}{llccc|ccc|ccc}
			& & \multicolumn{3}{c}{$(\eta,\sigma,\delta)=(2.5, 0.2, 1.5)$} &
			\multicolumn{3}{c}{$(\eta,\sigma,\delta)=(4, 0.2, 3)$} &\multicolumn{3}{c}{$(\eta,\sigma,\delta)=(4.5, 0.2, 3.5)$} \\
			\hline
			& Method & {$\hat{\text{bias}}^2$} & {$\hat{\text{var}}$} & {$\hat{\text{MSE}}$} 
			& {$\hat{\text{bias}}^2$} & {$\hat{\text{var}}$} & {$\hat{\text{MSE}}$}   & {$\hat{\text{bias}}^2$} & {$\hat{\text{var}}$} & {$\hat{\text{MSE}}$} \\
			\hline
   
			\multirow{4}{*}{} &
			RM    &1.9e-02&5.5e-02&7.3e-02
                &7.9e-03&1.3e-01&1.4e-01
			&4.2e-03&7.2e-02&7.6e-02\\
                
			&RM-DR     &7.2e-04 &\cellcolor{lightgreen}2.1e-03&\cellcolor{lightgreen}2.8e-03
                &\cellcolor{lightgreen}5.5e-04&\cellcolor{lightgreen}3.2e-03&\cellcolor{lightgreen}3.7e-03
                &\cellcolor{lightgreen}1.6e-04&\cellcolor{lightgreen}1.8e-03&\cellcolor{lightgreen}2.0e-03\\
            
			&SLE    &\cellcolor{lightgreen}6.2e-04&1.0e-02&1.1e-02
                &2.7e-03&9.1e-03&1.2e-02
			&4.3e-04&1.8e-02&1.9e-02 \\

                &ABC    &9.7e-03&3.0e-03&1.3e-02
                &3.5e-03&4.8e-03&8.3e-03
			&2.6e-03&7.2e-03&9.8e-03\\ 
                
			\hline	
		\end{tabular}
	\end{center}
 \normalsize
\end{table}

			
			
			

\subsection{M/G/1-queue}
We next consider a queuing model consisting of a first-come-first-serve single-server queue (M/G/1-queue), used in \citet{Fearnhead2012} as an example of a stochastic simulation model with an intractable likelihood. The service times are uniformly distributed on the interval $\left[\theta_1, \theta_2\right]$, and inter-arrival times are exponentially distributed with rate $\theta_3$. 
The simulation procedure for the $n$th inter-departure time $y(n)$ is provided in the supplement.
Assume that the first 1,000 inter-departure times $y = (y(1),\ldots, y(1,000))^\top$
are observed and the parameters $\theta=(\theta_1,\theta_2,\theta_3)^\top$ are unknown. 
Supplement Fig. S3 
shows a histogram of the inter-departure times from four independent realizations (panels) of $y$ when $\theta = \left(4, 8, 1/6\right)^\top$. 
The design distribution for $(\theta_1, \theta_2 - \theta_1, \theta_3)^\top$ is chosen as a uniform distribution over the region $\left(0, 10\right) \times \left(0, 10\right) \times \left(0, \frac{1}{3}\right)$, where the resulting service and inter-arrival times have on average comparable magnitudes.
%

The summaries chosen for implementation of ABC, SLE, and RM-DR provide information about the marginal distribution of inter-departure times: the minimum, maximum, and 18 evenly-spaced quantiles of $y$. This choice is motivated by exploratory analysis which suggests that the marginal distribution of $y$ may be more informative about $\theta$ than the time ordering. 

\begin{figure}
	\begin{center}
 		\setlength{\abovecaptionskip}{0pt} 
		\setlength{\belowcaptionskip}{0pt} 

		\subfigure{%
			
			\includegraphics[height=0.25\textwidth]{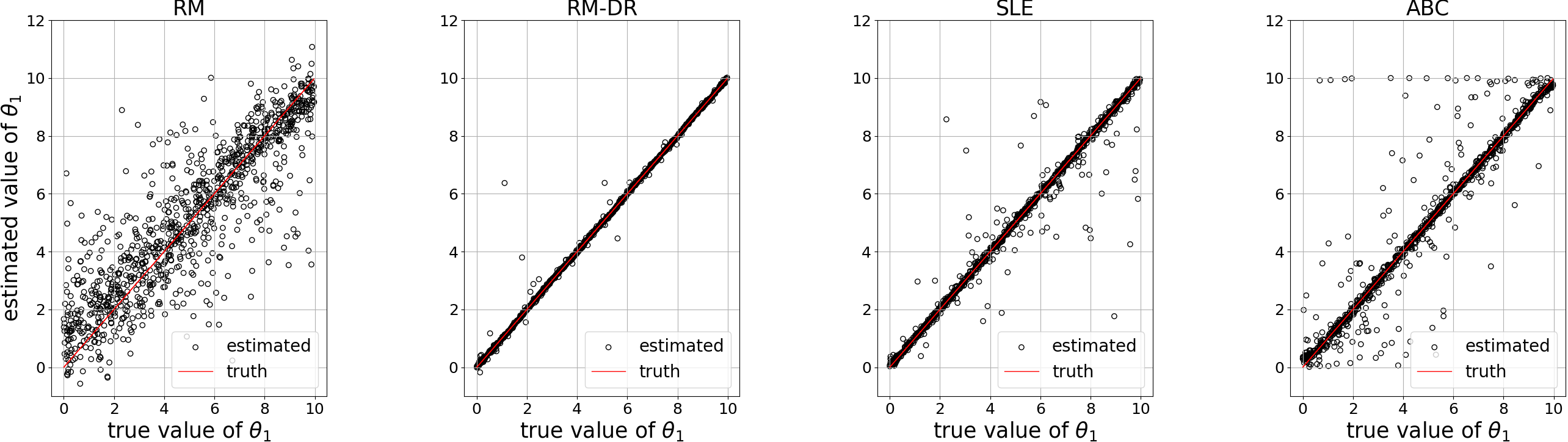}
		}\\ 
            \vspace{\subfigspacing}
		\subfigure{%
			
			\includegraphics[height=0.25\textwidth]{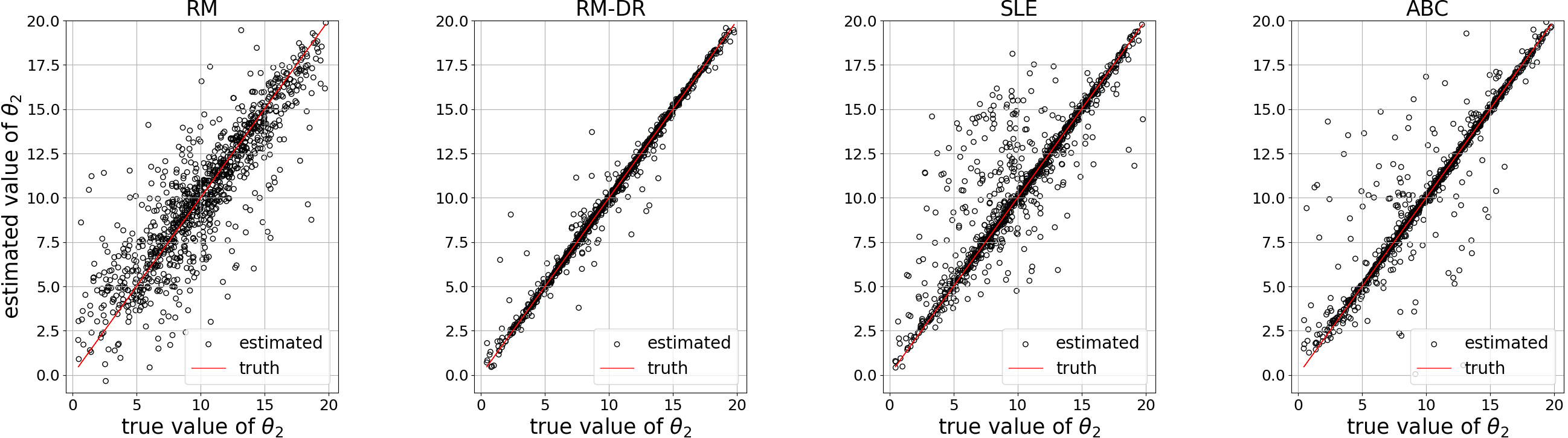}
		}\\ 
            \vspace{\subfigspacing}
		\subfigure{%
			
			\includegraphics[height=0.25\textwidth]{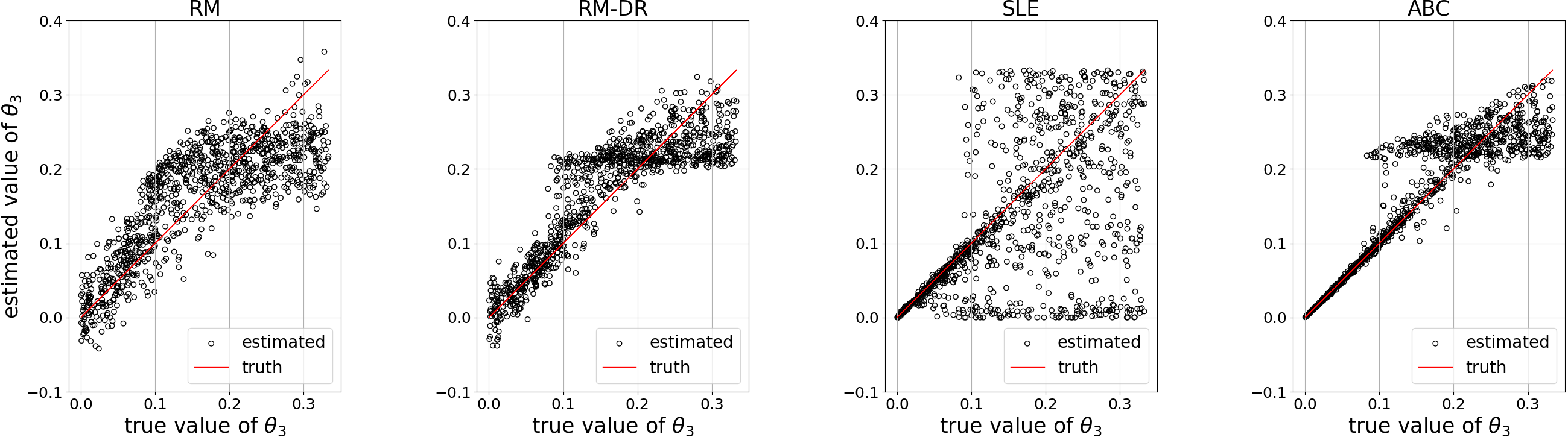}
		}
	\end{center}
        \vspace{-0.1in}
	\caption{Scatter plots of estimates versus simulation values of three components of $\theta$  (rows), respectively, using RM, RM-DR, SLE, and ABC (columns), respectively, for the M/G/1 model example. The 45\textdegree\ line is shown in red for reference.}
	\label{Scatter MG}
\end{figure}

Fig. \ref{Scatter MG} shows scatter plots of estimates versus simulation values of the components of $\theta$ (rows), under the four different estimation methods (columns). The smallest spread of values around the 45\textdegree\ line (red) indicating correct estimation is achieved by RM-DR. 
As in the previous example, supplement Fig. S4 
shows a comparison between the performance of RM and RM-DR estimators. 
On average, RM-DR achieves substantially better estimation performance than RM, in terms of squared bias, variance, and MSE across all the simulation parameter values. 
As in the previous example, SLE and ABC are further evaluated at three $\theta$ settings in Table \ref{table MG}. RM-DR has the best performance across all criteria, followed by ABC. The estimation performance of RM and SLE is substantially worse, with much higher values of squared bias and variance. Looking at MSE, RM performs marginally better than SLE, mainly due to lower estimation variance. 
Once again, the RM-DR method performs well in the likelihood-free setting under a sensible choice of summary statistics.  

\begin{table}
    \small
	\begin{center}
  \caption{MC approximation of squared bias, variance, and MSE for M/G/1 model example}
\label{table MG}
\begin{adjustbox}{max width=1.1\textwidth}
		\begin{tabular}{llccc|ccc|ccc}
			& & \multicolumn{3}{c}{$\theta=(9.502, 17.720, 0.244)$} &
			\multicolumn{3}{c}{$\theta=(8.119, 13.489, 0.092)$} &\multicolumn{3}{c}{$\theta=(9.594, 14.775, 0.309)$} \\
			\hline
			& Method & {$\hat{\text{bias}}^2$} & {$\hat{\text{var}}$} & {$\hat{\text{MSE}}$} 
			& {$\hat{\text{bias}}^2$} & {$\hat{\text{var}}$} & {$\hat{\text{MSE}}$}   & {$\hat{\text{bias}}^2$} & {$\hat{\text{var}}$} & {$\hat{\text{MSE}}$} \\
			\hline
			\multirow{4}{*}{} &
			RM    &2.4e-02&7.1e-01&7.4e-01
&4.6e-01&1.0&1.5
&3.0e-01&4.0e-01&7.0e-01\\
			
			&RM-DR     &\cellcolor{lightgreen}1.9e-03&\cellcolor{lightgreen}6.5e-03&\cellcolor{lightgreen}8.4e-03
&6.9e-03&\cellcolor{lightgreen}1.8e-02&\cellcolor{lightgreen}2.4e-02
&\cellcolor{lightgreen}1.0e-02&\cellcolor{lightgreen}1.9e-03&\cellcolor{lightgreen}1.2e-02\\

			&SLE    &4.0e-02&8.2e-01&8.6e-01
			&1.0e-01&1.9&2.0
                &3.0e-02&1.1&1.2\\ 
                
			&ABC    &2.0e-03&1.2e-02&1.4e-02
			&\cellcolor{lightgreen}4.1e-03&1.2e-01&1.2e-01
                &1.1e-02&5.0e-03&1.6e-02\\

			\hline	
		\end{tabular}
  \end{adjustbox}
	\end{center}
    \normalsize
\end{table}

			
			
			

\subsection{Lotka–Volterra Model}
Next, we consider estimation for the Lotka–Volterra (LV) model, used to describe the time evolution of abundance of two species in a prey-predator relationship. Key interactions between the two species can be captured by the three reaction types,
$$
u \longrightarrow 2 u, \quad
u+v \longrightarrow 2 v, \quad
v  \longrightarrow \emptyset,
$$
where $u$ and $v$ represent the abundance of a prey and predator species, respectively. The first reaction describes prey production (e.g., through birth or immigration), the second reaction captures consumption of prey by the predator, and the third reaction represents removal of predators (e.g. through death or out-migration). These dynamics can be described by a continuous-time discrete state Markov chain, where each reaction occurs at a rate that depends on the current state of the system, specified in terms of transition probabilities over a small time interval $\left(t, t+\delta t\right]$, as explained in the supplement. 
We denote the state of the system at time $t$ by $y(t)=\left(u(t), v(t)\right)^\top$, where $u(t)$ and $v(t)$ represent the abundance of prey and predators at time $t$, respectively. 
Assume the initial condition $y(0)=(50,100)^\top$ and unknown parameters $\theta=(\theta_1,\theta_2,\theta_3)^\top \in (0.3,0.6)\times (0.005,0.01)\times (0.1,0.4)$. In this example, we observe both prey and predator populations  at 1,000 equidistant points in a time interval $\left[0,30\right]$, and denote the observed abundances by $u\in \mathbb{R}^{1,000}$ and $v\in \mathbb{R}^{1,000}$, respectively. Given $\theta$, we simulate data using the Gillespie algorithm \citep{Gillespie1977},  
as illustrated in supplement Fig. S5.  
%

Although the RM approach could in principle be generalized to multivariate data, the method as originally proposed uses univariate observations. Therefore, for the RM implementation we concatenate the two vectors as $y=(u^\top,v^\top)^\top$ as inputs. For the remaining estimation approaches, we utilize the same summaries for both predator and prey variables. 
The statistics we consider include those used in the Ricker model example based on similar justifications, in addition to 20 B-spline regression coefficients, and sample cross-correlation to capture the temporal structure of the data and the relationship between $u$ or $v$.

\begin{figure}
	\begin{center}
		\subfigure{%
			
			\includegraphics[height=0.25\textwidth]{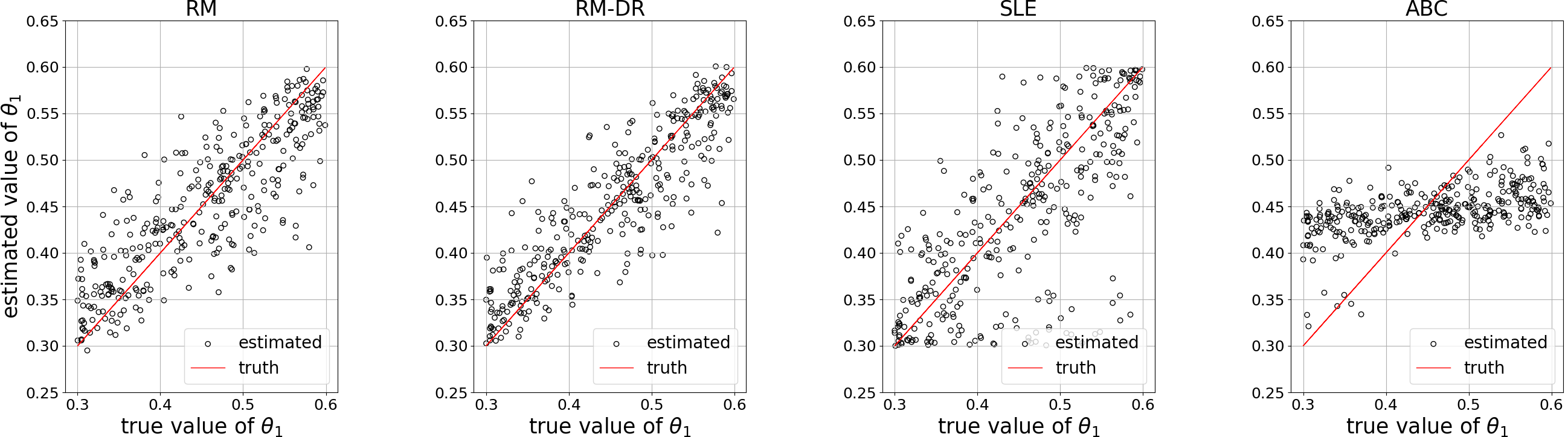}
		}\\ 
            \vspace{\subfigspacing}
		\subfigure{%
			
			\includegraphics[height=0.25\textwidth]{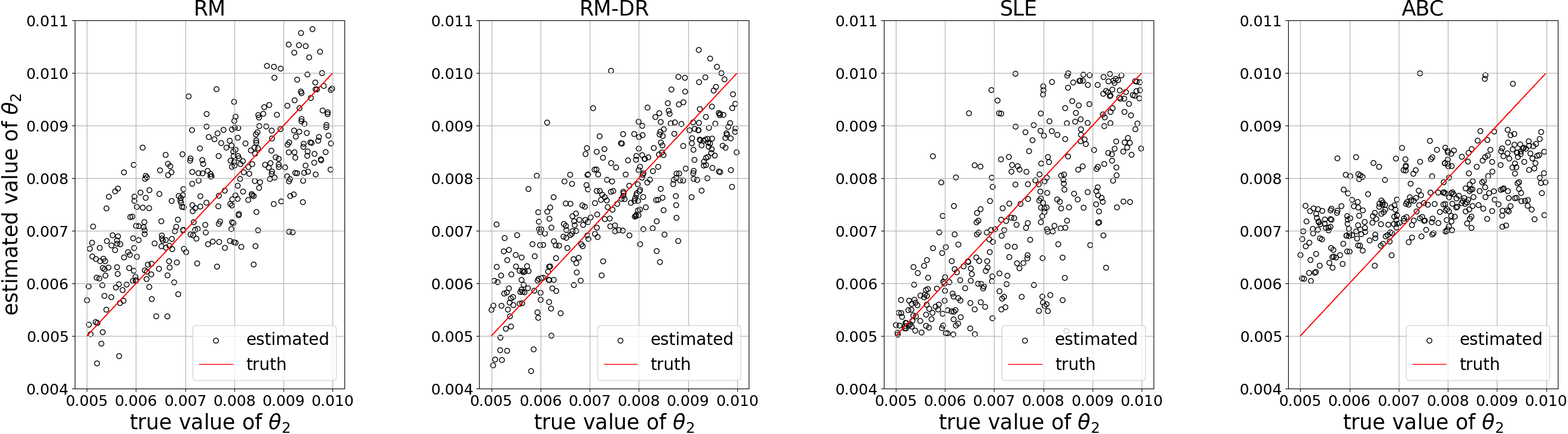}
		}\\ 
            \vspace{\subfigspacing}
		\subfigure{%
			
			\includegraphics[height=0.25\textwidth]{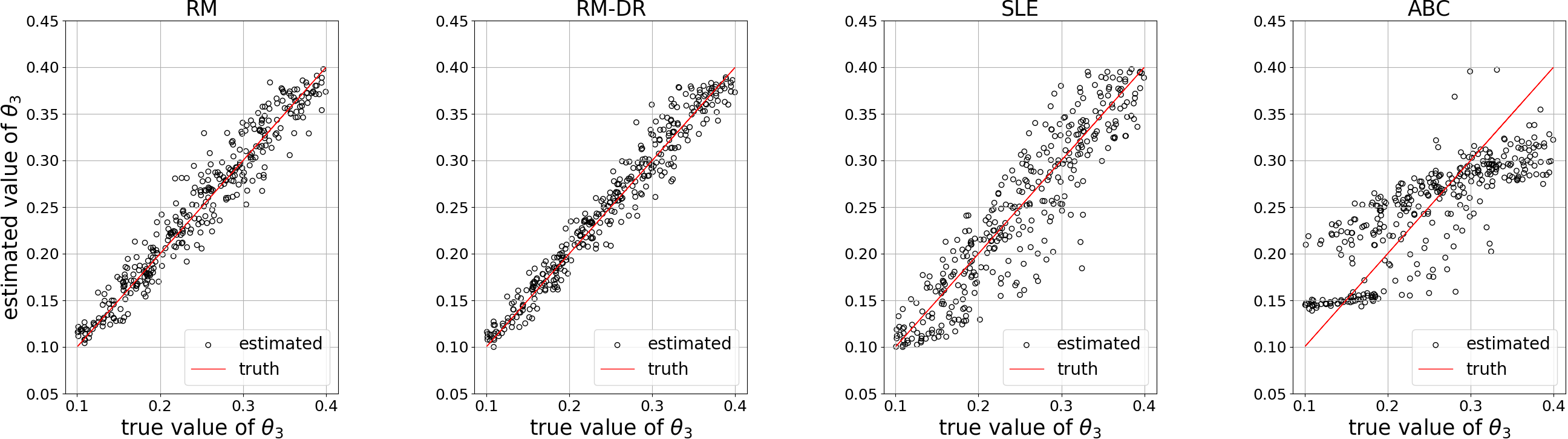}
		}
	\end{center}
        \vspace{-0.1in}
	\caption{Scatter plots of estimates versus simulation values of three components of $\theta$  (rows), respectively, using RM, RM-DR, SLE, and ABC (columns), respectively, for the LV model example. The 45\textdegree\ line is shown in red for reference.}
	\label{Scatter LV}
\end{figure}

Looking at the scatter plots of estimated values versus simulation values of parameters in Fig. \ref{Scatter LV}, RM-DR estimates are both more accurate and less variable than RM and SLE. While ABC has low variability overall, it does have high bias. Performance of RM and RM-DR is illustrated in supplement Fig. S6 
at 200 uniformly sampled $\theta$ settings. RM-DR estimation has lower squared bias, variance and MSE for most $\theta$ values relative to RM estimation. It also has lower integrated versions of these metrics, which demonstrates a better overall performance. Again we evaluate performance of ABC and SLE at several $\theta$ settings in Table \ref{table LV}, showing that RM-DR has the lowest MSE and squared bias across the estimation methods considered. ABC is second to RM-DR in terms of MSE, mainly due to having low variance, while SLE has the worst overall performance. 

\begin{table}
        \small
	\begin{center}
  \caption{MC approximation of squared bias, variance and MSE for LV model example}
\label{table LV}
\begin{adjustbox}{max width=1.1\textwidth}
		\begin{tabular}{llccc|ccc|ccc}
			& & \multicolumn{3}{c}{$\theta=(0.59,0.0077,0.392)$} &
			\multicolumn{3}{c}{$\theta=(0.431,0.0051,0.265)$} &\multicolumn{3}{c}{$\theta=(0.465,0.0085,0.187)$} \\
			\hline
			& Method & {$\hat{\text{bias}}^2$} & {$\hat{\text{var}}$} & {$\hat{\text{MSE}}$} 
			& {$\hat{\text{bias}}^2$} & {$\hat{\text{var}}$} & {$\hat{\text{MSE}}$}   & {$\hat{\text{bias}}^2$} & {$\hat{\text{var}}$} & {$\hat{\text{MSE}}$} \\
			\hline
			\multirow{4}{*}{} &
			RM    &2.9e-03&1.2e-03&4.1e-03
&2.7e-03&1.6e-03&4.3e-03
&5.4e-04&2.2e-03&2.8e-03\\
			
			&RM-DR     &\cellcolor{lightgreen}1.1e-03&5.8e-04&\cellcolor{lightgreen}1.7e-03
&\cellcolor{lightgreen}7.7e-04&9.9e-04&\cellcolor{lightgreen}1.8e-03
&\cellcolor{lightgreen}3.4e-04&1.9e-03&\cellcolor{lightgreen}2.3e-03\\

			&SLE    &1.8e-02&9.0e-03&2.7e-02
			&2.0e-03&2.4e-03&4.4e-03
                &1.8e-03&7.9e-03&9.7e-03\\

               &ABC    &2.2e-02&\cellcolor{lightgreen}4.1e-04&2.2e-02
			&2.1e-03&\cellcolor{lightgreen}3.0e-04&2.4e-03
                &1.0e-03&\cellcolor{lightgreen}1.5e-03&2.6e-03\\ 
   
			\hline	
		\end{tabular}
  \end{adjustbox}
	\end{center}
        \normalsize
\end{table}

			
			
			

\subsection{FitzHugh–Nagumo Model}
The final numerical example considers parameter estimation for the FitzHugh–Nagumo (FN) ODE model, which describes the time evolution of voltage $v(t)$ and recovery $r(t)$ across the membrane of a biological neuron. The ODE initial value problem (see supplement) depends on unknown parameters $\theta=(\theta_1,\theta_2)^\top$, fixed constants $\tau = 3$ and $\zeta = 0.4$, and initial conditions $v(0)=r(0)=0$.
%
%
We make the standard assumption that only voltage is observed with additive noise via $y(t) =v(t) + \epsilon$ 
at a discrete set of locations $t_i=0.025i$ for $i=1,\ldots, 1,000$, and with $\epsilon \stackrel{\text{i.i.d.}}{\sim} \mathcal{N}(0,0.06^2)$. 
The ODE solution $v(t)$ under different parameter settings is shown in supplement Fig. S7.  
Since the likelihood for this model can be approximated numerically, we can compare RM and RM-DR estimation with maximum likelihood estimation. The summaries chosen for RM-DR implementation are the coefficients of a nonlinear regression on $K$ Fourier basis functions, chosen as a way of extracting frequency and amplitude information from this periodic system. 

\begin{figure}
	\begin{center}
        \includegraphics[height=0.3\textwidth]{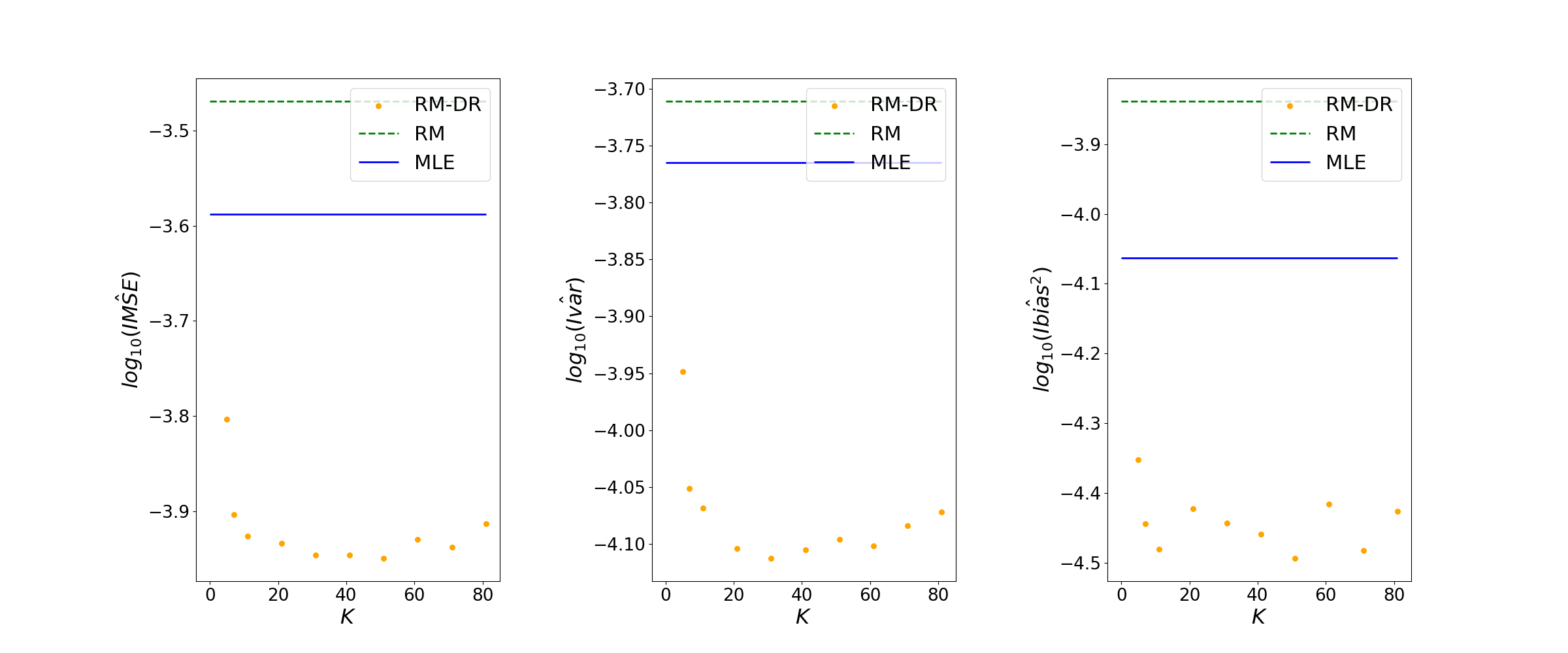}
	\end{center}
        \vspace{-0.2in}
	\caption{Plots of MC approximation of integrated squared bias, variance and MSE in FN model, $K$ is the dimension of input space in RM-DR method.}
	\label{FN_integrated}
\end{figure}

For the integrated performance metrics, we consider test parameter values over the grid $(\theta_1, \theta_2) \in \{-0.2+0.03 j \, , \, -0.4+0.04 l\}_{j, l = 0,1,\ldots,40}$. We vary the number of basis functions $K$ to investigate how the summary dimension impacts RM-DR performance. Fig. \ref{FN_integrated} shows that RM-DR has the smallest integrated squared bias, variance, and MSE of the three methods, regardless of input dimension. Its performance is robust to different choices of input dimension within a reasonable range for this example. As expected, using $K=5$ basis coefficients leads to notably worse performance than using larger values of $K$, as the latter choices convey more amplitude and frequency information. Unsurprisingly, RM has the largest integrated squared bias and variance due to the difficulty in reconstructing a mapping with a large input space. 
%
Supplement Fig. S8 shows Monte Carlo estimates of log squared bias, variance, and MSE for the three methods considered. For all approaches, estimation is worse for simulation parameter values in the top left triangular region of the parameter space. This is because the ODE solution associated with these parameter quickly attains a steady state, as shown in supplement Fig. S7, containing less information about the parameter.
RM-DR provides an improvement over RM at many parameter values. Comparing RM-DR with MLE, RM-DR has better estimations in the top left region of the parameter space. Although MLE produces lower bias and better estimation at most parameter values, it has larger variance in the top left region, which leads to a higher IMSE relative to RM-DR. Because the likelihood is available in this example, we can also test the RM-DRLO approach described in the supplement, which consists of using RM-DR as a starting for to a local optimization algorithm as an alternative to an expensive global optimization method. The results are described in supplement section 4.4.

\section{Summary}
We propose a simulation-based RM-DR estimation approach with dimension reduction for the class of inverse problems in which a closed-form likelihood is unavailable or expensive, and discuss its properties, evaluation criteria, and uncertainty quantification. This approach resolves the problem of degraded estimation performance when data dimension increases, which makes direct reconstruction map estimation unreliable in practice. We show that under mild assumptions, the RM-DR estimator converges to a Bayes estimator in probability. By learning a dimension-reduced manifold, RM-DR reduces the approximation error relative to RM estimation, as illustrated in multiple numerical experiments. Additionally, in the setting where the likelihood is available but expensive, we propose to combine the RM-DR approach with local optimization methods as an alternative to global optimization approaches for parameter estimation, with comparable accuracy but more time efficiency. 

In the numerical examples, RM-DR stands out as a highly effective approach for parameter estimation in complex models with intractable likelihoods, demonstrating clear advantages over other popular methods in terms of accuracy and computational efficiency, especially when estimation for multiple datasets under the same model is of interest. By leveraging informative summary statistics and reducing the dimensionality of the input space, RM-DR effectively captures the important features of the data, reducing estimation error and leading to more accurate parameter estimates compared to the RM method. Unlike the ABC approach, which rejects training data that are in some sense far away from the observed data, RM-DR essentially utilizes all the parameter-data pairs in the construction of the estimator. When trained on a sufficiently large number of synthetic samples, this results in robust and adaptable estimations. Notably, RM-DR outperforms SLE, which relies on the strong assumption of normality in the synthetic likelihood, making RM-DR a more robust and widely applicable choice for complex modeling scenarios.

In terms of future work, for high-dimensional data where informative application-specific summaries are not readily available—particularly in complex or less interpretable settings—we propose exploring the automatic learning of summary statistics. This could involve using unsupervised learning techniques, such as autoencoders, to reduce the dimensionality of the data before inputting it into the neural network model. Alternatively, we may incorporate a transformer encoder directly into the neural network architecture to summarize key features in sequential data. By leveraging the transformer's self-attention mechanism to capture long-range dependencies and contextual relationships, the model may learn richer representations, potentially improving the robustness and accuracy of parameter estimation.


\section*{Data Availability Statement}
All numerical experiments were based on a large number of simulated datasets. No real data was analyzed as part of this work. 

\bibliography{mref}

\end{document}


\bibliographystyle{chicago}

\if0\blind
{
  \title{\bf Supplement for ``Dimension-reduced Reconstruction Map Learning for Parameter Estimation in Likelihood-Free Inference Problems''}
  \author{Rui Zhang\\ 
    Department of Statistics, The Ohio State University\\
    and \\
    Oksana Chkrebtii \\
    Department of Statistics, The Ohio State University\\
    and \\
    Dongbin Xiu \\
    Department of Mathematics, The Ohio State University
    }
  \maketitle
} \fi

\if1\blind
{
  \bigskip
  \bigskip
  \bigskip
  \begin{center}
    {\LARGE\bf Title}
\end{center}
  \medskip
} \fi

\newpage
\spacingset{1.75} 

\section{Additional Background}

\subsection{Likelihood-based Methods}
When the likelihood function is either analytically or computationally tractable, standard likelihood-based estimation methods can be applied. \textit{Maximum likelihood estimation} (MLE) takes the estimator to be,
\begin{equation}
    \hat{\theta}_{MLE}=\underset{\theta\in \Theta}{\operatorname{argmax}} \; \log p(y\mid \theta), \label{MLE}
\end{equation}
that is, the parameters under which the observed data has the highest relative frequency. A popular alternative to maximum likelihood estimation is the \textit{penalized likelihood estimator},
\begin{equation}
        \hat{\theta}_{PLE}=\underset{\theta\in \Theta}{\operatorname{argmax}} \;\log(p(y\mid \theta))-\lambda q(\theta),\label{PLE}
\end{equation}
which adjusts the objective function by subtracting a penalty term from the log likelihood. The tuning parameter $\lambda$ controls the tradeoff between the fit to the data and the penalty, and $q(\theta)$ is a non-negative function that can encode prior knowledge about parameters or penalize what we think of as unrealistic estimates. 
A common example is a ridge penalty term with $q(\theta)=\norm{\theta}_2^2$, where $\norm{x}_p$ denotes $L^p$ norm of $x$ throughout this paper. This choice of penalty can shrink the estimates towards zero, since $q(\theta)$ is large for parameters with large magnitude. 
A \textit{Bayes estimator} is a function,
\begin{equation}
        \hat{\theta}_{B}=\underset{\hat{\theta}}{\operatorname{argmin}}\;  r(\pi,\hat{\theta}),\label{Bayes}
\end{equation}
that minimizes the \textit{Bayes risk} $r(\pi,\hat{\theta}):=\mathbb{E}_{(y,\theta)}\bigl[l(\theta,\hat{\theta}(y))\bigr]$ over all possible functions of $y$, where the expectation is taken with respect to the joint distribution with density $p(y,\theta)=\pi(\theta)p(y\mid \theta)$ and $\pi(\theta)$ is a prior density. 
When the likelihood is not available computationally or in closed form, another class of methods is needed.

\subsection{Simulation-based Methods}\label{Formulation}

Simulation-based inference methods do not require likelihood evaluation, and are applicable when fast simulation from the generative model is possible at arbitrary parameter values. 
Popular approaches include \textit{approximate Bayesian computation} (ABC) and \textit{synthetic likelihood estimation} (SLE). 

The principle underlying ABC is that given a prior density $\pi(\cdot)$ over $\theta$, the posterior density can be written as,
\begin{equation}
       \pi(\theta\mid y)\propto \pi(\theta)\int p(y'\mid \theta)I\left(y', y\right) dy',\label{posterior}
\end{equation}
where $I$ is the indicator function. When the likelihood $p(y'\mid \theta)$ cannot be evaluated, a Monte Carlo approximation $\frac{1}{N_s}\sum_{i = 1}^{N_s} I\left(y_{i}, y\right)$ of the integral part in (\ref{posterior}) can, in principle, be constructed using synthetic data $y_{1}, \ldots, y_{N_s} \sim p(y\mid \theta)$ drawn from the generative model.
In practice, this approach is not feasible for continuous data, where the probability of generating a sample equal to $y$ is zero, and is potentially inaccurate for discrete data of high dimension due to large Monte Carlo error resulting from the low probability of generating an exact match to the full data. Approximate Bayesian inference is based on a Markov chain Monte Carlo or sequential Monte Carlo sample targeting an approximation of (\ref{posterior}) where the indicator is replaced by a kernel function which puts higher weight on synthetic data that is close to $y$ based on an appropriate distance metric. To further reduce Monte Carlo error, a function $S:\mathbb{R}^{m}\rightarrow \mathbb{R}^{K}$ is chosen to summarize the $m$-dimensional data by a $K$-dimensional statistic. Producing close matches of the full data simultaneously becomes increasingly unlikely as $m$ grows, so a choice of $K<m$ can reduce Monte Carlo error while introducing further approximation when $S$ is a not sufficient for $\theta$. 
We denote by $s=S(y)$ the summary of the observed data $y$. The resulting \textit{ABC posterior} takes the form,
\begin{equation}
       \pi_{\text{ABC}}(\theta\mid s)\propto \pi(\theta)\int p(y'\mid \theta)K\left[\{S(y')-s\}/h\right]dy',\label{ABC posterior}
\end{equation}
where $K(\cdot)$ is a kernel function and $h>0$ is a bandwidth parameter. An ABC estimate of $\theta$ can be taken as functional of the ABC posterior, such as the ABC posterior mean, 
\begin{equation}
       \hat{\theta}_{ABC}=\mathbb{E}_{\text{ABC}}\bigl [\theta\mid s\bigr].\label{ABC}
\end{equation}
Selection of the bandwidth $h$ provides a trade-off between accuracy and Monte Carlo error. As $h$ tends to zero, the kernel $K$ tends to the indicator of an exact match between the simulated and summarized data. This results in difficulty accepting a large enough number of simulated samples, thereby increasing Monte Carlo error. Tuning the bandwidth parameter is an application-dependent problem and depends on both the structure of the data, the number of summaries, and the available computing resources relative to the complexity of simulating from the model.

Similarly, SLE relies on summarization of simulated data $y_{1}, \ldots, y_{N_s} \sim p(y\mid \theta)$. It works by numerically maximizing a synthetic likelihood of $s_{i}=S(y_{i})$ obtained via a multivariate normal approximation to the distribution of the sample summaries $s_{1}, \ldots, s_{N_s}$. For a given $\theta$ the log synthetic likelihood is,
\begin{equation}
  l_s(\theta)=-\frac{1}{2}(s-\hat{\mu}_{\theta})^\top \hat{\Sigma}_{\theta}^{-1}(s-\hat{\mu}_{\theta})-\frac{1}{2}\log \mid \hat{\Sigma}_{\theta}\mid , \label{synthetic likelihood} 
\end{equation}
where $\hat{\mu}_{\theta}=\sum_{i=1}^{N_s}s_{i}/N_s, \hat{\Sigma}_{\theta}=\sum_{i=1}^{N_s} (s_{i}-\hat{\mu}_{\theta})(s_{i}-\hat{\mu}_{\theta})^\top /N_s$, and the maximum synthetic likelihood estimator is,
\begin{equation}
    \hat{\theta}_{SLE}=\underset{\theta\in \Theta}{\operatorname{argmax}} \; l_s(\theta). \label{synthetic likelihood estimator} 
\end{equation}

Another simulation based method, which we will call \textit{reconstruction map estimation}, has recently been proposed by \citep{Rudi2021} and is introduced in Section 3.1 of the main paper.  

\subsection{Fitting Neural Networks to Data}

Given a fixed NN architecture, to fit a NN model to training data $\mathcal{H}$, the values of the NN parameters must be optimized by optimizing an \textit{objective function} $L(\omega)$, which quantifies estimation performance. The objective function is usually chosen to be the \textit{training loss}, defined as $L(\omega)=\frac{1}{|\mathcal{H}|}\sum\limits_{x\in \mathcal{H}}l(x\mid \omega)$, where $l(x\mid \omega)$ denotes the loss for one training sample $x$ under NN parameters $\omega$. Since the optimization problem does not have a closed-form solution, numerical optimization via \textit{gradient descent} is typically used. The iteration step of the algorithm is 
\begin{equation}
        \omega_{j}=\omega_{j-1}-\alpha \Delta L(\omega_{j-1}),  \label{gradient descent}
\end{equation}
where $\Delta L(\omega_{j-1})$ is the gradient of $L(\omega_{j-1})$, and $\alpha$ is the \textit{learning rate}, the rate at which algorithm updates parameters. Since computation of the gradient based on the full data at each iteration is expensive, a \textit{mini-batch gradient descent} algorithm is widely used. It partitions the entire training data into $b$ batches $\mathcal{B}_1,\dots,\mathcal{B}_b$, and it uses one batch of the data to approximate the gradient in each iteration step 
\begin{equation}
        \omega_{j}=\omega_{j-1}-\alpha \frac{1}{|\mathcal{B}_{a_j}|}\sum\limits_{x\in \mathcal{B}_{a_j}}\Delta l(x\mid \omega_{j-1}),  \label{mini-batch gradient descent}
\end{equation}
where $a_j$ indexes the batch chosen at iteration $j$, $\{a_j\}$ is a periodic sequence with period $b$, and $a_j=j \text{, for } j=1,\dots, b$ without loss of generality. So every $b$ iteration steps, all batches are fed exactly once to train the model and update the parameters, and this procedure is referred to as one \textit{epoch}. Additionally, it has been shown in studies that fixed learning rates often produce sub-optimal performance \citep{Duchi2011, Bengio2012}, and therefore it is necessary to let the learning rate gradually decay as the algorithm proceeds. Adaptive learning rate methods based on incorporating a notion of momentum \citep{Rumelhart1986} have also been proposed, including AdaGrad\citep{Duchi2011}, RMSprop \citep{Hinton2012}, Adam \citep{Kingma2015}. In this paper we use mini-batch gradient descent with Adam as our default optimization algorithm.

\section{Proofs}

\subsection{Proof of Theorem 1}

\begin{proof}
Let \((\Xi, \mathcal{F}, P)\) denote the underlying probability space with respect to which all convergence notions are defined, where \(\Xi\) is the sample space (set of all outcomes), \(\mathcal{F}\) is a $\sigma$-algebra of measurable subsets of \(\Xi\), and \(P\) is a probability measure on \(\mathcal{F}\).

To prove convergence in probability, it suffices to show that any subsequence of $\{\hat{\omega}_n\}$ has a further subsequence along which the corresponding functions converge pointwise almost surely to $\mathbf{N}_0(\cdot)$. This implies convergence pointwise in probability of the entire sequence $\mathbf{N}(\cdot, \hat{\omega}_n)$ to $\mathbf{N}_0(\cdot)$ as $n\to\infty$.

Let \( \{\hat{\omega}_{n_k}\} \) be an arbitrary subsequence of \( \{\hat{\omega}_n\} \). Since each \( \hat{\omega}_{n_k} \) takes values in the compact set \( \Omega \), by compactness there exists a further subsequence \( \{\hat{\omega}_{m_j}\} \) and a random variable \( \omega^* \) such that:
\[
\hat{\omega}_{m_j} \xrightarrow{a.s.} \omega^*, \text{ as } j\to \infty.
\]
Define $\Xi_1 := \left\{ \xi \in \Xi : \lim_{j \to \infty}\hat{\omega}_{m_j}(\xi) = \omega^*(\xi) \right\},$
so that \( P(\Xi_1) = 1 \). Next, by the uniform convergence assumption,
\[
\sup_{\omega \in \Omega} |Q_n(\omega) - Q_0(\omega)| \xrightarrow{p} 0 , \text{ as } n\to \infty, 
\]
and by the subsequence principle, there exists a further subsequence of \( \{m_j\} \) (which we continue to denote by \( \{m_j\} \) for notational simplicity) such that:
\[
\sup_{\omega \in \Omega} |Q_{m_j}(\omega) - Q_0(\omega)| \xrightarrow{a.s.} 0, \text{ as } j\to \infty.
\]
Denote \( Q_n(\omega, \xi) \) as the realization of the random function \( Q_n(\omega) \) at the outcome \( \xi \in \Xi \). Define $
\Xi_2 := \left\{ \xi \in \Xi :  \lim_{j \to \infty}\sup_{\omega \in \Omega} |Q_{m_j}(\omega, \xi) - Q_0(\omega)| = 0 \right\},$ so that $ P(\Xi_2) = 1$. And we let 
$\Xi' := \Xi_1 \cap \Xi_2,$ so that $P(\Xi') = 1.$

Now, for any $\xi \in \Xi'$, by definition of $\hat{\omega}_{m_j}(\xi)$ as a minimizer,
$Q_{m_j}(\hat{\omega}_{m_j}(\xi), \xi) \leq Q_{m_j}(\omega_0, \xi)$ for any $\omega_0 \in \Omega_0$. Since \(\lim_{j \to \infty}\sup_{\omega \in \Omega} |Q_{m_j}(\omega, \xi) - Q_0(\omega)| =0\) for all \(\xi \in \Xi'\), it follows that for any fixed \(\omega \in \Omega\), \(\lim_{j \to \infty}Q_{m_j}(\omega, \xi) = Q_0(\omega)\). Moreover, since \(\lim_{j \to \infty}\hat{\omega}_{m_j}(\xi) =\omega^*(\xi)\), and the convergence of \(Q_{m_j}(\omega, \xi)\) to \(Q_0(\omega)\) is uniform in \(\omega\), we can conclude that \(\lim_{j \to \infty}Q_{m_j}(\hat{\omega}_{m_j}(\xi), \xi) = Q_0(\omega^*(\xi))\). On the other hand, since \(\omega_0\) is fixed, we have \(\lim_{j \to \infty}Q_{m_j}(\omega_0, \xi) = Q_0(\omega_0)\). Because the inequality \(Q_{m_j}(\hat{\omega}_{m_j}(\xi), \xi) \leq Q_{m_j}(\omega_0, \xi)\) holds for every \(j\) and both sides converge,  we pass to the limit in the inequality, yielding \(Q_0(\omega^*(\xi)) \leq Q_0(\omega_0)\) for all \(\xi \in \Xi'\).

Since $\omega_0$ is a minimizer of $Q_0$, the inequality implies:
    \[
    Q_0(\omega^*(\xi)) = Q_0(\omega_0),
    \]
    showing that $\omega^*(\xi) \in \Omega_0$ for all $\xi \in \Xi'$. By the unique minimizing NN function assumption, any $\omega \in \Omega_0$ induces the same function
$\mathbf{N}(\cdot, \omega) = \mathbf{N}_0(\cdot). $
Therefore, we conclude that
\begin{equation}\label{NN minimization function}
    \mathbf{N}(\cdot, \omega^*(\xi)) = \mathbf{N}_0(\cdot)
\end{equation}
for all \(\xi \in \Xi'\), which holds almost surely. Finally, by the continuity of \(\mathbf{N}(\cdot, \omega)\) in \(\omega\) and the almost sure convergence \(\hat{\omega}_{m_j} \xrightarrow{a.s.} \omega^*\), we have that for each fixed $s\in \mathcal{S}$, \(\mathbf{N}(s, \hat{\omega}_{m_j}) \xrightarrow{a.s.} \mathbf{N}(s, \omega^*)\) as $j \to \infty$. Since we have established that \(\mathbf{N}(\cdot, \omega^*) = \mathbf{N}_0(\cdot)\) almost surely, we conclude that for each fixed $s\in \mathcal{S}$,
\[
\mathbf{N}(s, \hat{\omega}_{m_j}) \xrightarrow{a.s.} \mathbf{N}_0(s), \text{ as } j \to \infty.
\]

Since the original subsequence \(\{\hat{\omega}_{n_k}\}\) was arbitrary, we have shown that every subsequence of \(\{\hat{\omega}_n\}\) admits a further subsequence along which the corresponding functions converge pointwise almost surely to \(\mathbf{N}_0(\cdot)\). This implies that the entire sequence \(\mathbf{N}(\cdot, \hat{\omega}_n)\) converges pointwise in probability to \(\mathbf{N}_0(\cdot)\) as $n\to\infty$, that is, for each fixed $s\in \mathcal{S}$,
\[
\hat{\theta}_n(s) \xrightarrow{p} \mathbf{N}_0(s), \text{ as } n\to\infty.
\]
This completes the proof of pointwise convergence in probability. 

Next, under the additional assumptions that the support of the summary statistics \(\mathcal{S}\) is compact and that \(\mathbf{N}(s, \omega)\) is jointly continuous on \(\mathcal{S} \times \Omega\), we prove uniform convergence in probability, by following the same initial steps as in the proof of pointwise convergence. We follow all steps to establish that for any subsequence $\{ \hat{\omega}_{n_k} \}$, there exists a further subsequence $\{ \hat{\omega}_{m_j} \}$ and a random variable $\omega^*$ such that:
\begin{enumerate}
    \item $\hat{\omega}_{m_j} \xrightarrow{a.s.} \omega^*$ as $j \to \infty$.
    \item $\omega^*$ is a minimizer of the expected training loss, so $\omega^* \in \Omega_0$ almost surely.
    \item $\mathbf{N}(\cdot, \omega^*) = \mathbf{N}_0(\cdot)$ almost surely.
\end{enumerate}

It is known that a continuous function on a compact domain is uniformly continuous. Since $\mathbf{N}(s, \omega)$ is jointly continuous on the compact set $\mathcal{S} \times \Omega$, it is also uniformly continuous on this domain. This uniform continuity implies that for any $\epsilon > 0$, there exists a $\delta > 0$ such that for any $\omega_a, \omega_b \in \Omega$:
\begin{equation}\label{uniform continuity property}
    \text{if } \|\omega_a - \omega_b\| < \delta, \quad \text{then} \quad \sup_{s \in \mathcal{S}} |\mathbf{N}(s, \omega_a) - \mathbf{N}(s, \omega_b)| < \epsilon.
\end{equation}
For any $\xi \in \Xi'$ (the set of probability 1 where the convergence holds), we have already established that $\lim_{j \to \infty} \hat{\omega}_{m_j}(\xi) = \omega^*(\xi)$. Thus there exists a $J$ such that for all $j > J$, we have $\|\hat{\omega}_{m_j}(\xi) - \omega^*(\xi)\| < \delta$. From (\ref{NN minimization function}) and (\ref{uniform continuity property}), it follows that for all $j > J$, $$\sup_{s \in \mathcal{S}} |\mathbf{N}(s, \hat{\omega}_{m_j}(\xi)) - \mathbf{N}_0(s)| < \epsilon.$$ 
Since this holds for any $\epsilon > 0$, we conclude that $\lim_{j \to \infty}\sup_{s \in \mathcal{S}} |\mathbf{N}(s, \hat{\omega}_{m_j}(\xi)) - \mathbf{N}_0(s)|=0$ for any $\xi \in \Xi'$. Thus,
\[
\sup_{s \in \mathcal{S}} |\mathbf{N}(s, \hat{\omega}_{m_j}) - \mathbf{N}_0(s)| \xrightarrow{a.s.} 0, \text{ as } j\to\infty.
\]
Since the original subsequence $\{\hat{\omega}_{n_k}\}$ was arbitrary, we have shown that every subsequence of $\{\hat{\theta}_n(\cdot)\}$ admits a further subsequence that converges uniformly almost surely to $\mathbf{N}_0(\cdot)$. By the subsequence principle, it follows that the entire sequence converges uniformly in probability. That is:
\[
\sup_{s \in \mathcal{S}} |\hat{\theta}_n(s) - \mathbf{N}_0(s)| \xrightarrow{p} 0, \text{ as } n \to \infty.
\]
This completes the proof of uniform convergence in probability under the strengthened assumptions.
\end{proof}

\subsection{Proof of Theorem 2}
\begin{proof}
Since $d(\theta)$ and $\pi(\theta)$ agree except on a set of measure zero, for any $\omega \in \Omega$, we have $$\mathbb{E}_{(s,\theta)\sim p_{d}(s,\theta)}\bigl[l(\theta, \mathbf{N}(s,\omega))\bigr] = \mathbb{E}_{(s,\theta)\sim p_{\pi}(s,\theta)}\bigl[l(\theta, \mathbf{N}(s,\omega))\bigr] = r_{s}(\pi, \mathbf{N}(\cdot,\omega)).$$ 
By Theorem 1, under Assumptions 1--4, there exists $\mathbf{N}_0(\cdot)$ such that for each fixed $s\in \mathcal{S}$, $\hat{\theta}_n(s) = \mathbf{N}(s,\hat{\omega}_n) \xrightarrow{p} \mathbf{N}_0(s)$, and $r_{s}(\pi, \mathbf{N}_0(\cdot)) \leq r_{s}(\pi, \mathbf{N}(\cdot,\omega))$ for all $\omega \in \Omega$. Since $\hat{\theta}_{B}(\cdot) \in \mathcal{A}$, we have $r_{s}(\pi, \mathbf{N}_0(\cdot)) \leq r_{s}(\pi, \hat{\theta}_{B}(\cdot))$. By the definition of the Bayes estimator, $r_{s}(\pi, \hat{\theta}_{B}(\cdot)) \leq r_{s}(\pi, \mathbf{N}_0(\cdot))$, and thus $r_{s}(\pi, \mathbf{N}_0(\cdot)) = r_{s}(\pi, \hat{\theta}_{B}(\cdot))$. By the uniqueness of $\mathbf{N}_0(\cdot)$ under Assumption 3, it follows that $\mathbf{N}_0(\cdot) = \hat{\theta}_{B}(\cdot)$, and thus for each fixed $s\in \mathcal{S}$, $$\hat{\theta}_n(s) \xrightarrow{p} \hat{\theta}_{B}(s), \text{ as } n\to\infty.$$
This completes the proof of pointwise convergence in probability to the Bayes estimator. 

Moreover, if additional Assumptions 5--6 in Theorem 1 hold, we have $\sup_{s \in \mathcal{S}} |\hat{\theta}_n(s) - \mathbf{N}_0(s)| \xrightarrow{p} 0, \text{ as } n \to \infty$. Since $\mathbf{N}_0(\cdot) = \hat{\theta}_{B}(\cdot)$, it follows that $$\sup_{s \in \mathcal{S}} |\hat{\theta}_n(s) - \hat{\theta}_{B}(s)| \xrightarrow{p} 0, \text{ as } n \to \infty.$$
This completes the proof of uniform convergence in probability to the Bayes estimator under the strengthened assumptions.
\end{proof}

\section{Additional Contributions}

\subsection{Uncertainty Quantification for RM-DR estimators}
So far, we have focused on estimation in the likelihood-free setting. We now turn to the problem of uncertainty quantification for RM and RM-DR estimators by constructing bootstrap confidence intervals. An approach for approximating the sampling distribution over $\theta$ is by using the parametric Bootstrap. Let $\hat{\theta}: \mathbb{R}^{m} \rightarrow \mathbb{R}^{d}$ be either the RM or RM-DR estimator. Using the estimate  $\hat{\theta}=\hat{\theta}(y)$, we generate B Bootstrap samples $y_1,\ldots,y_B$ by sampling from the data-generating distribution
\begin{equation}
    y_b\stackrel{\text{ind}}{\sim}p(y\mid \hat{\theta}), \quad b = 1,\ldots, B.
\end{equation}
For each Bootstrap sample $y_b$, we compute the Bootstrap estimate $\hat{\theta}_b=\hat{\theta}(y_b)$. The collection of Bootstrap estimates $\{\hat{\theta}_b\}_{b=1,\ldots,B}$ is then used to obtain the empirical Bootstrap sampling distribution of the estimator and any desired probability intervals. For instance, we can construct confidence intervals for any component of $\theta$ by using percentiles of Bootstrap estimates. 
Suppose $\theta_{[j]}$ is the $j$th component of the parameter vector, then a $100(1-\alpha) \%$ Bootstrap confidence interval for $\theta_{[j]}$ is 
\begin{equation}
    \left(\hat{\theta}_{[j]}^{\;\frac{\alpha}{2}}, \hat{\theta}_{[j]}^{\; 1-\frac{\alpha}{2}}\right), \label{Bootstrap CI}
\end{equation}
where $\hat{\theta}_{[j]}^{\;\frac{\alpha}{2}}$ and $\hat{\theta}_{[j]}^{\; 1-\frac{\alpha}{2}}$ are the $100 \frac{\alpha}{2}$ and $100(1-\frac{\alpha}{2})$ percentiles of $\{\hat{\theta}_{b,[j]}\}_{b=1, \ldots,B}$. 

A Bootstrap confidence region for $\theta$ is constructed similarly based on the Bootstrap sample mean $\overline{\hat{\theta}\,}=\sum_{b=1}^{B}\hat{\theta}_b/B$ and sample covariance matrix $\hat\Sigma=\frac{1}{B-1}\sum_{b=1}^{B}({\hat\theta}_b-\overline{{\hat\theta}\,})({\hat\theta}_b-\overline{{\hat\theta}\,})^\top$. The $100(1-\alpha)\%$ confidence region for $\theta$ can be approximated as
\begin{equation}
    \{\theta: (\overline{\hat{\theta}\,}-\theta)^\top \hat{{\Sigma}}^{-1} (\overline{\hat{\theta}\,}-\theta) \leq \chi^2_d(1-\alpha)\},
\end{equation}
where $\chi^2_d(1-\alpha)$ is the $100(1-\alpha)$ percentile of a chi-squared distribution with $d$ degrees of freedom. 

While the proposed parametric Bootstrap approach is straightforward to implement, and can be applied as long as the generative model is known, it is also computationally intensive as it requires simulating a large number of Bootstrap samples from the model to conduct inference.  


\subsection{A Combined RM-DR and Local Optimization (RM-DRLO) Method}\label{sec:rmdrlo}

So far, we have considered settings where the likelihood is not accessible. However, inference is sometimes challenging when the likelihood is available but expensive to evaluate enough times to use global optimization, while being relatively inexpensive to sample from. This is where RM-DR estimation can be combined with local optimization to speed up estimation. After obtaining the RM-DR estimate by evaluating data on a fitted reconstruction map, we provide it as the starting point to a less expensive local optimization approach targeting the objective function, in this case the log-likelihood. We take the result of the local optimization as the estimate when a local convergence criterion is met. Examples of local optimization methods include Nelder-Mead, Broyden–Fletcher–Goldfarb–Shanno (BFGS), Newton's method and so on. We call this as combined RM-DR and local optimization (RM-DRLO) approach. Without loss of generality, assume the update rule for a given local algorithm in the iteration step is $\theta^{j}=\Psi(\theta^{j-1})$, and $\Psi^n(\theta)=\underbrace{f \circ f\circ \cdots \circ f }_n(\theta)$ is an iterated function that applies the update rule $n$ times. Suppose the algorithm is run for $N_l$ total number of iterations that is based on a stopping criterion. The RM-DRLO estimator is
\begin{equation}
    \hat{\theta}_{RM-DRLO}(y)=\Psi^{N_l}(\theta_0(y)), \text{ where }\theta_0(y)=\hat{\theta}_{RMDR}(S(y)).  \label{RM-DRLO estimator}
\end{equation}
By using RM-DR estimation, we narrow down the search space and find a solution that is sub-optimal, then a local optimization algorithm is better able to find the most optimal solution within that region with respect to the desired objective function. In subsequent sections, numerical experiments illustrate how this combined estimation approach achieves comparable performance to the estimation provided by a global optimization method that minimizes the same cost function, while being much more computationally efficient.

\section{Additional Results and Details for Numerical Experiments}

\subsection{Ricker model}

\begin{figure}[H]
	\begin{center}
	\includegraphics[height=0.4\textwidth]{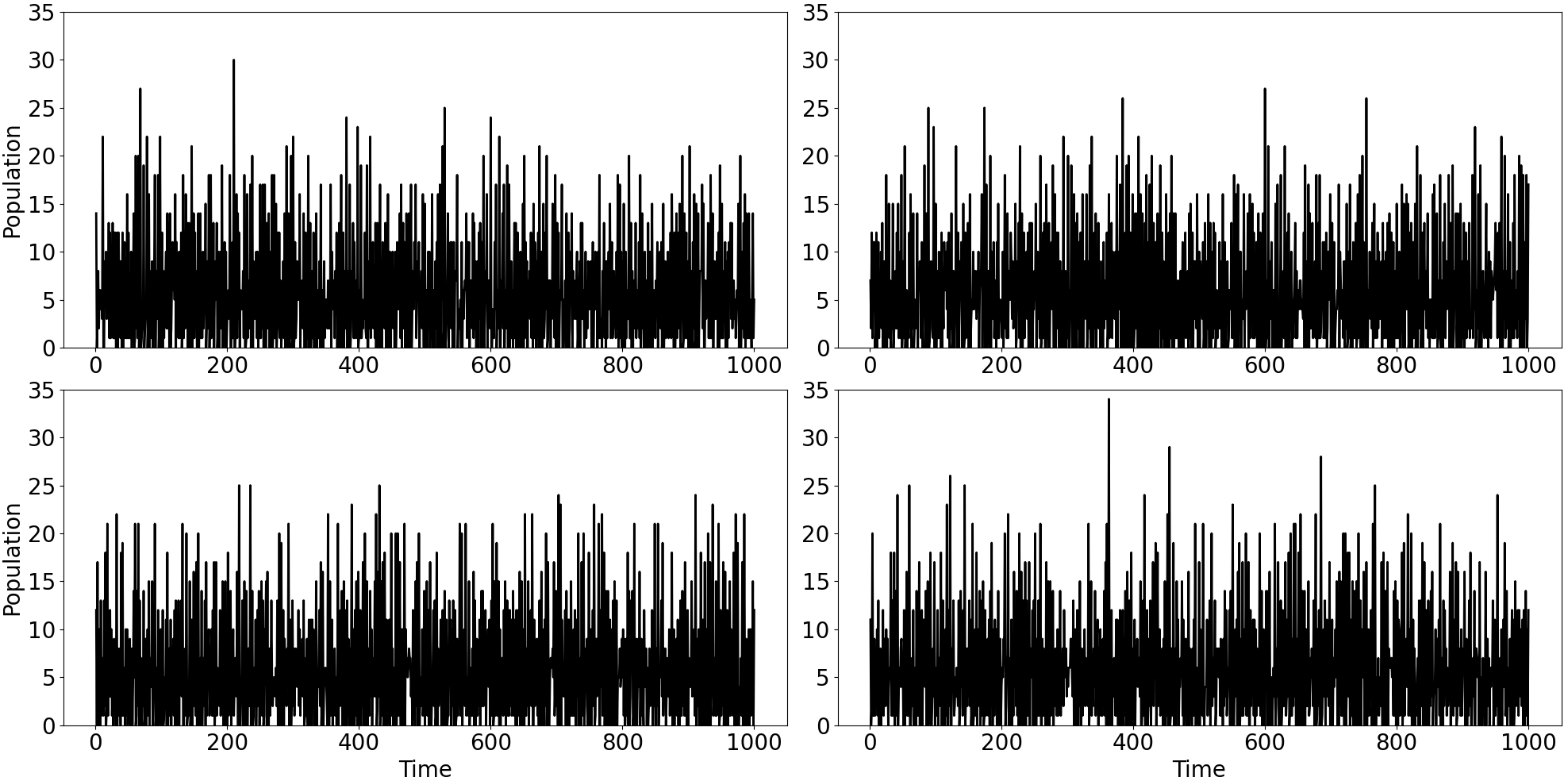}
	\end{center}
	\caption{Four realizations of observed animal population simulated from (\ref{Poisson observation}) under the Ricker model with parameters $\eta=3, \sigma=0.2,\delta=2$.}
	\label{fig1}
\end{figure}

The population density $N(t)$ is updated across a set of discrete time steps $t \in \mathbb{Z}^+$ via,
\begin{equation}
  N(t+1)=aN(t)e^{-N(t)+\epsilon(t)}, \label{Ricker model} 
\end{equation}
where $\epsilon(t)\stackrel{\text{ind}}{\sim}\mathcal{N}(0,\sigma^2)$ represents process noise within the dynamical system, and $a$ is an intrinsic growth rate parameter. We model the observed population size using Poisson model with mean $\delta N(t)$
\begin{equation}
  y(t)\stackrel{\text{ind}}{\sim} \text{Poisson}(\delta N(t)), \label{Poisson observation} 
\end{equation}
where $\delta$ is an unknown scale parameter. The initial population is set to $N(0)=2$, and data at $m = 1,000$ consecutive time steps, $y=(y(1),\ldots,y(1,000))^\top$, is observed. Setting $\eta=\log(a)$, the parameters of interest are $\theta=(\eta,\sigma,\delta)^\top$, and the target parameter space is $\Theta=(2,5)\times(0,0.3)\times(1,4)$. Figure \ref{fig1} shows four replications of $y$ simulated under the parameter setting $\theta = (3, 0.2, 2)^\top$.  A likelihood calculation would require marginalization over $m$ unobserved population densities, and is thus effectively intractable. 

\begin{figure}[H]
	\begin{center}
		\subfigure{%
			
			\includegraphics[width=0.6\textwidth,height=0.2\textwidth]{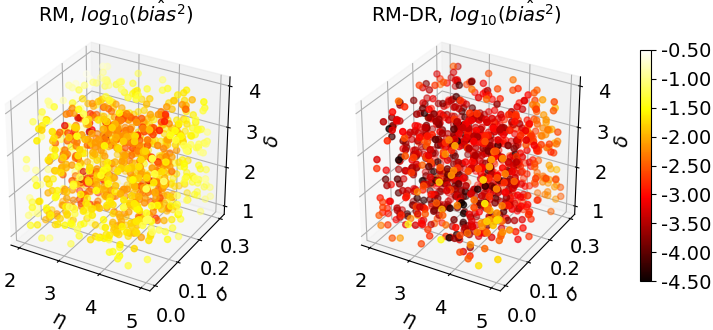}
		}\\
		\subfigure{%
			
			\includegraphics[width=0.6\textwidth,height=0.2\textwidth]{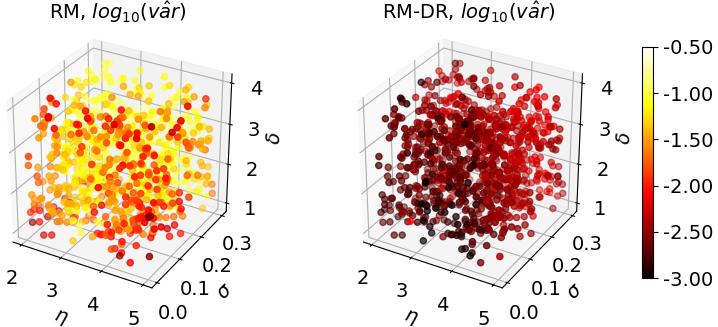}
		}\\
		\subfigure{%
			
			\includegraphics[width=0.6\textwidth,height=0.2\textwidth]{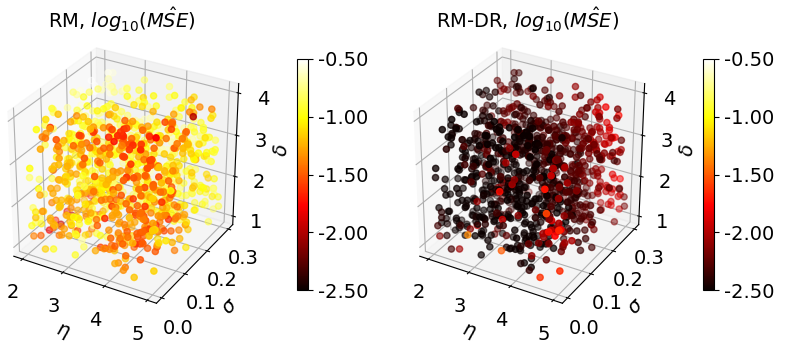}
		}
	\end{center}
	\caption{Magnitude (color) of the log squared bias, variance, and MSE (along rows) for RM (left column) and RM-DR (right column), respectively, under different parameter settings (points in 3-d space) for the Ricker model example. MC estimates of integrated performance criteria are $\hat{\text{IBIAS}}^2$=2.9e-02, $\hat{\text{IVAR}}$=7.1e-02, $\hat{\text{IMSE}}$=1.0e-01 for RM, and $\hat{\text{IBIAS}}^2$=1.5e-03, $\hat{\text{IVAR}}$=3.4e-03, $\hat{\text{IMSE}}$=4.9e-03 for RM-DR.}
	\label{Heatmap Ricker}
\end{figure}

\subsection{M/G/1-queue}

 Let $u(n)$ be the service time for the $n$th customer, and $w(n)$ be the difference between the arrival time of the $n$th and the $(n-1)$th customer, with $w(1)=0$. Inter-departure times $y(n)$ (the difference between the departure time of the $n$th and $(n-1)$th customer, with $y(1)=u(1)$) are generated as
$$
y(n)=\left\{\begin{array}{l}
u(n), \quad \text { if } \sum_{i=1}^n w(i) \leq \sum_{i=1}^{n-1} y(i) \\
u(n)+\sum_{i=1}^n w(i)-\sum_{i=1}^{n-1} y(i), \quad \text { otherwise. }
\end{array}\right.
$$

\begin{figure}[H]
	\begin{center}
	\includegraphics[height=0.25\textwidth]{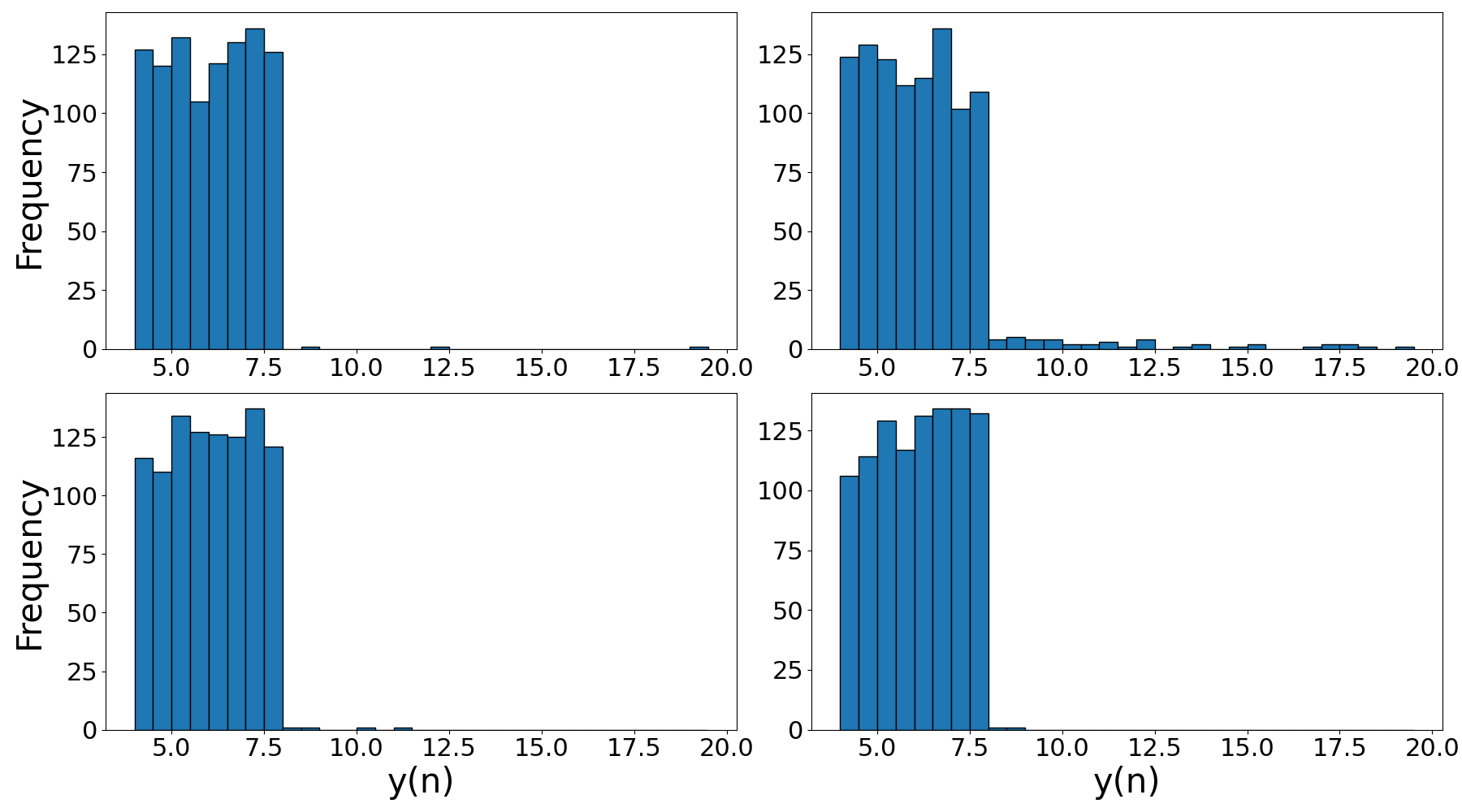}
	\end{center}
	\caption{Histogram of inter-departure times from four independent realizations (panels) of the M/G/1-queue with parameters $\theta_1=4, \theta_2=8, \theta_3=\frac{1}{6}$.}
	\label{fig2}
\end{figure} 

\begin{figure}[H]
	\begin{center}
		\subfigure{%
			
			\includegraphics[width=0.6\textwidth,height=0.2\textwidth]{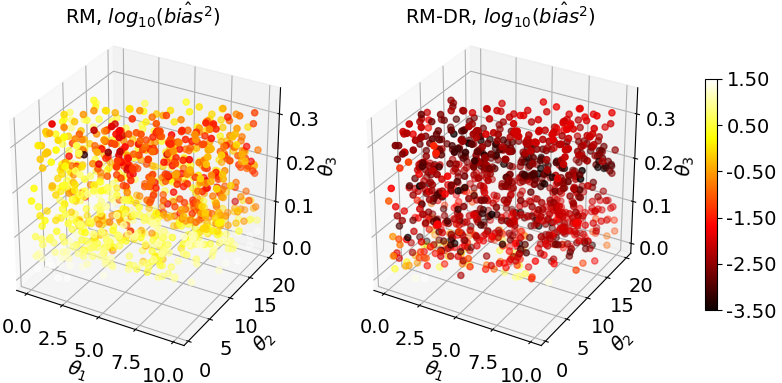}
		}\\
		\subfigure{%
			
			\includegraphics[width=0.6\textwidth,height=0.2\textwidth]{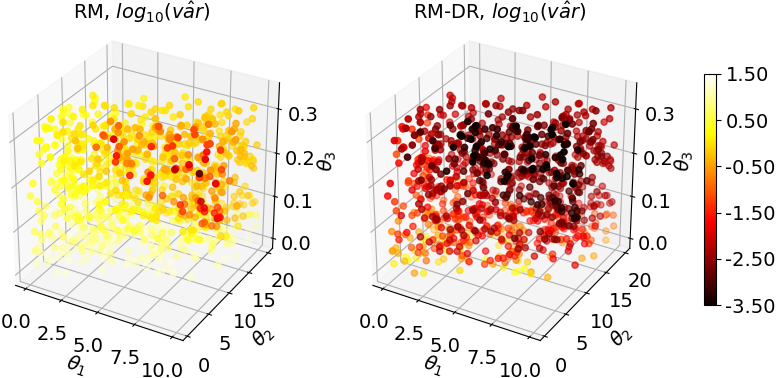}
		}\\
		\subfigure{%
			
			\includegraphics[width=0.6\textwidth,height=0.2\textwidth]{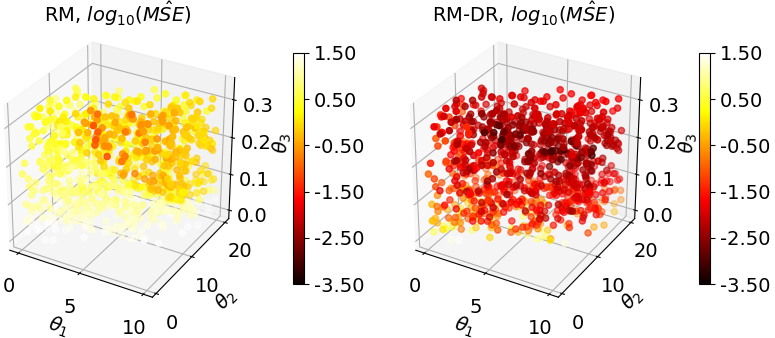}
		}
	\end{center}
	\caption{Magnitude (color) of the log squared bias, variance, and MSE (along rows) for RM (left column) and RM-DR (right column), respectively, under different parameter settings (points in 3-d space) for the  M/G/1 model  example. MC estimates of integrated performance criteria are  $\hat{\text{IBIAS}}^2$=3.4, $\hat{\text{IVAR}}$=4.0, $\hat{\text{IMSE}}$=7.5 for RM, and $\hat{\text{IBIAS}}^2$=2.2e-01, $\hat{\text{IVAR}}$=8.7e-02, $\hat{\text{IMSE}}$=3.1e-01 for RM-DR.}
	\label{Heatmap MG}
\end{figure}

\subsection{Lotka–Volterra model}

The dynamics of the Lotka-Volterra model can be described by a continuous-time discrete state Markov chain, where each reaction occurs at a particular rate that depends on the current state of the system. Formally, it can be specified in terms of transition probabilities over a small time interval $\left(t, t+\delta t\right]$. We denote the state of the system at time $t$ as $y(t)=\left(u(t), v(t)\right)^\top$, where $u(t)$ and $v(t)$ represent the abundance of prey and predators at time $t$ in the population, respectively. The transition probabilities are
\begin{equation}\label{nice}
\begin{split}
\Pr\{y(t+\delta t)&=\left(u^{*}, v^{*}\right)^\top \mid y(t)=\left(u, v\right)^\top\}\\&=
    \begin{cases}
        1-\left(\theta_1 u+\theta_2 u v+\theta_3 v\right) \delta t+o(\delta t), & \text { if } u^{*}=u, v^{*}=v \\
        \theta_1 u \delta t+o(\delta t), & \text { if } u^{*}=u+1, v^{*}=v \\
        \theta_2 u v \delta t+o(\delta t), & \text { if } u^{*}=u-1, v^{*}=v+1 \\
        \theta_3 v \delta t+o(\delta t) & \text { if } u^{*}=u, v^{*}=v-1 \\
o(\delta t), & \text { otherwise,}\end{cases}
\end{split}
\end{equation}
for $t \geq 0$ and small positive $\delta t$. Here $\theta_1$ represents the reproduction rate of prey, $\theta_2$ is the consumption rate of prey by the predator, and $\theta_3$ denotes the removal rate of the predator. 
\begin{figure}[H]
	\begin{center}
	\includegraphics[height=0.35\textwidth]{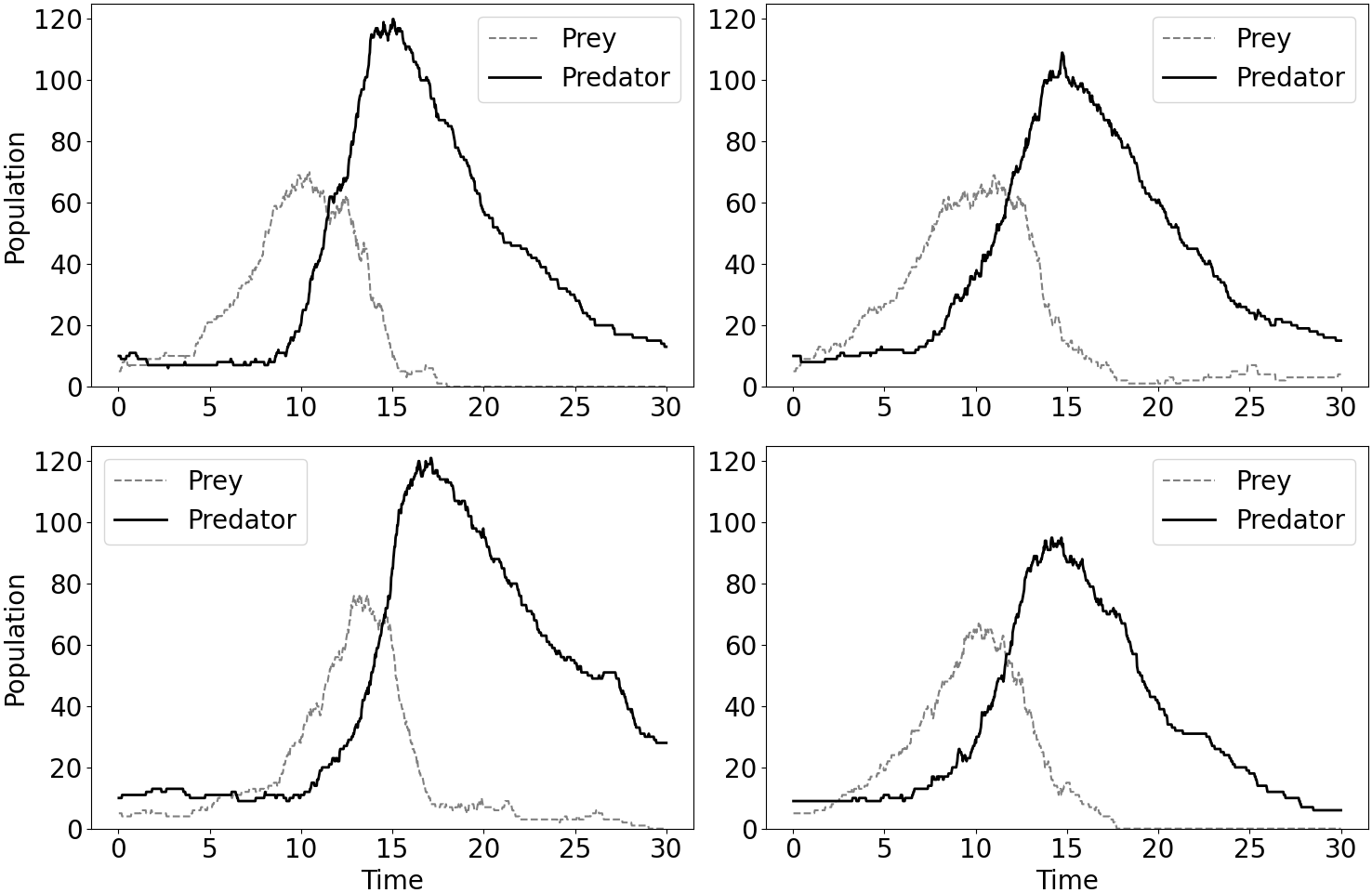}
	\end{center}
	\caption{Population of prey and predator simulated from the LV model with parameter value $\theta_1=0.35, \theta_2=0.009, \theta_3=0.15$}
	\label{fig3}
\end{figure} 

\begin{figure}[H]
	\begin{center}
		\subfigure{%
			
			\includegraphics[width=0.6\textwidth,height=0.2\textwidth]{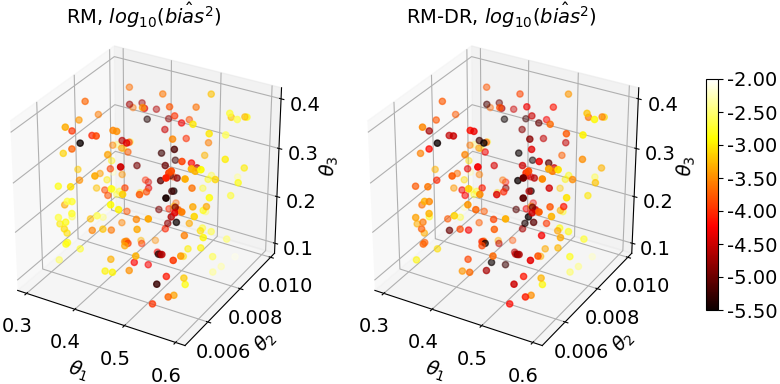}
		}\\
		\subfigure{%
			
			\includegraphics[width=0.6\textwidth,height=0.2\textwidth]{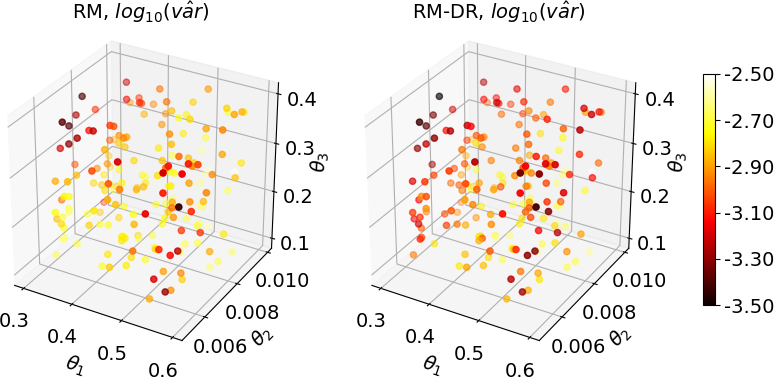}
		}\\
		\subfigure{%
			
			\includegraphics[width=0.6\textwidth,height=0.2\textwidth]{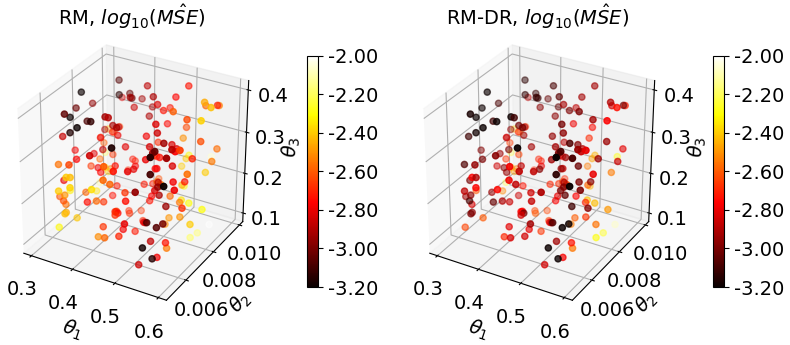}
		}
	\end{center}
 \caption{Magnitude (color) of the log squared bias, variance, and MSE (along rows) for RM (left column) and RM-DR (right column), respectively, under different parameter settings (points in 3-d space) for the LV model example. MC estimates of integrated performance criteria are  $\hat{\text{IBIAS}}^2$=8.1e-04, $\hat{\text{IVAR}}$=1.5e-03, $\hat{\text{IMSE}}$=2.3e-03. For RM-DR, $\hat{\text{IBIAS}}^2$=4.5e-04, $\hat{\text{IVAR}}$=1.2e-03, $\hat{\text{IMSE}}$=1.6e-03.}
	\label{Heatmap LV}
\end{figure}

\subsection{FitzHugh–Nagumo Model}

The governing equations for the membrane voltage $v(t)$ and recovery $r(t)$ at time $t$ are,
\begin{equation}
  \left\{
    \begin{aligned}
       \frac{\mathrm{d} v }{\mathrm{d} t} &=\tau\left(v-\frac{v^{3}}{3}+r+\zeta\right) \\
       \frac{\mathrm{d} r }{\mathrm{d} t} &=-\frac{1}{\tau}\left(v-\theta_{1}+\theta_{2} r\right) 
    \end{aligned}
  \right. ,\label{FN model} 
\end{equation}
with initial conditions $v(0)=r(0)=0$, unknown parameters $\theta=(\theta_1,\theta_2)^\top$, and fixed constants $\tau = 3$ and $\zeta = 0.4$.

\begin{figure}[H]
	\begin{center}
	\includegraphics[height=0.2\textwidth]{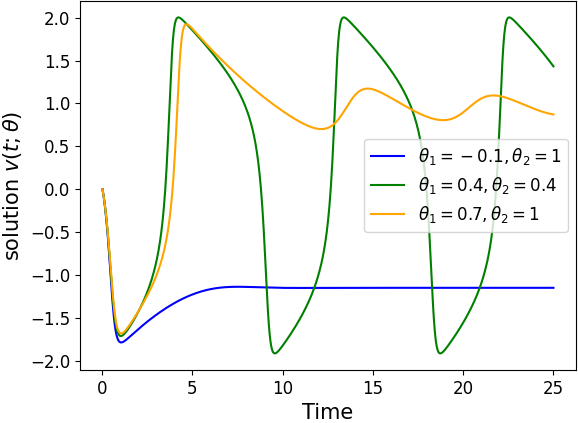}
	\end{center}
        \vspace{-0.2in}
	\caption{Marginal solution $v(t)$ of the FN model under three different $\theta$ settings (legend).}
	\label{FN_traj}
\end{figure}


\begin{figure}[H]
	\begin{center}
		\subfigure{%
			
			\includegraphics[width=0.7\textwidth,height=0.3\textwidth]{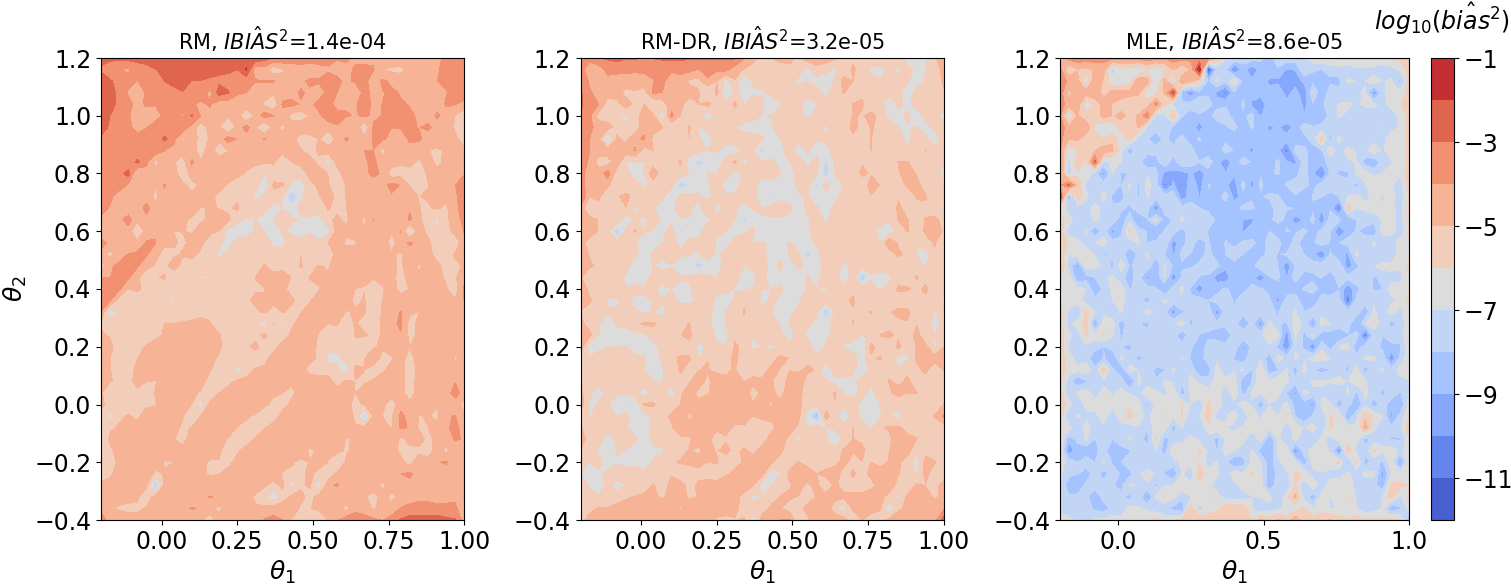}
		}\\
		\subfigure{%
			
			\includegraphics[width=0.7\textwidth,height=0.3\textwidth]{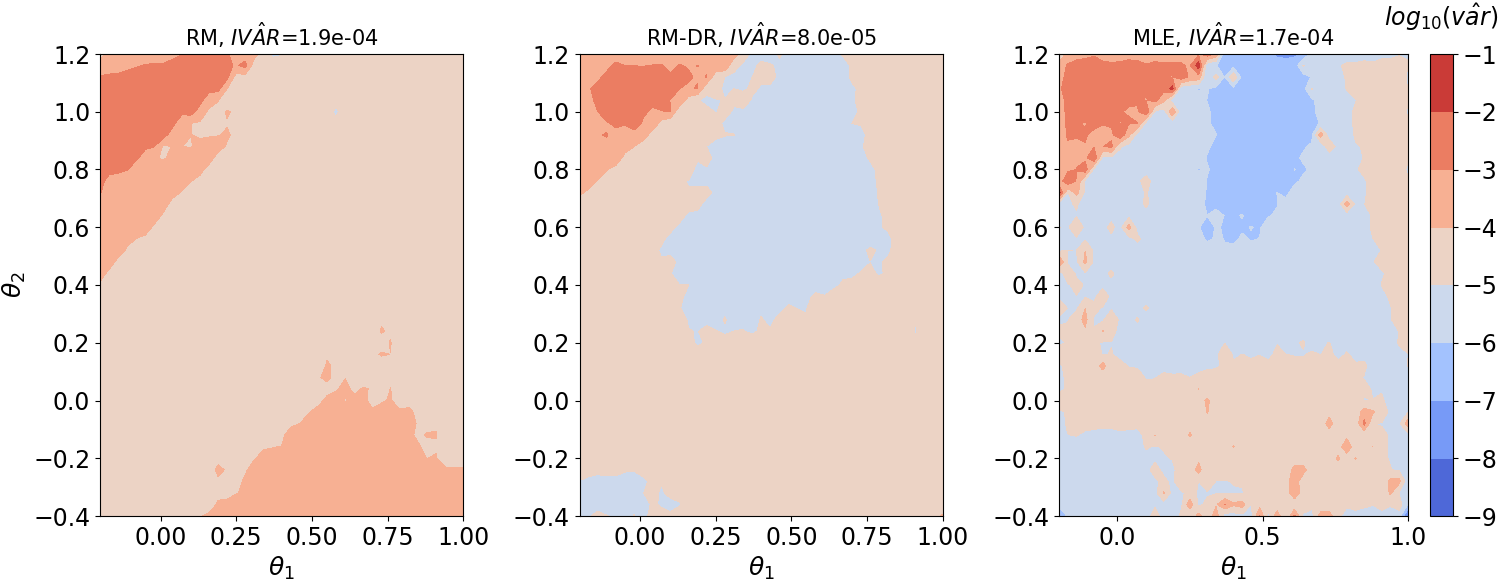}
		}\\
		\subfigure{%
			
			\includegraphics[width=0.7\textwidth,height=0.3\textwidth]{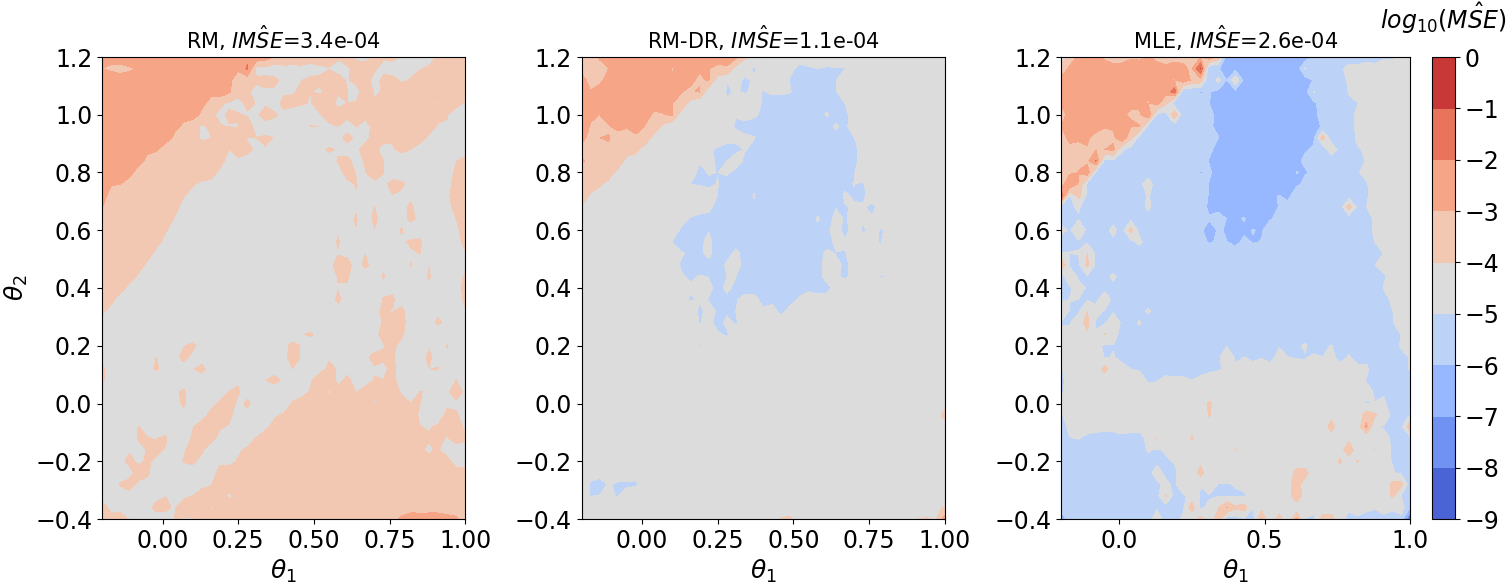}
		}
	\end{center}
        \vspace{-0.2in}
	\caption{Monte Carlo estimates of log squared bias, variance and MSE, respectively (colormap), for the FN model. Results of RM-DR estimation are based on summarization with $K=51$ Fourier series regression coefficients.}
	\label{FN_metrics}
\end{figure}

\subsubsection{Testing the RM-DRLO method}

\begin{figure}[H]
	\begin{center}
	\includegraphics[height=0.3\textwidth]{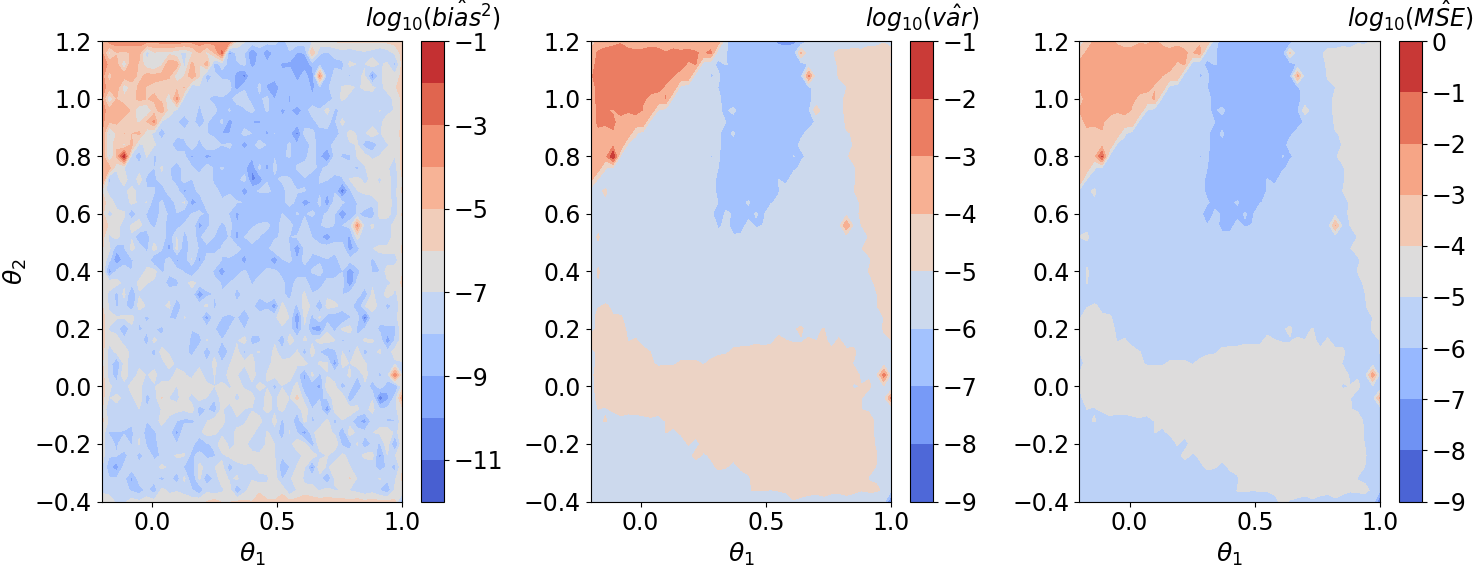}
	\end{center}
        \vspace{-0.15in}
	\caption{Monte Carlo estimates of log squared bias, variance, and MSE for the RM-DRLO method. $\hat{\text{IBIAS}}^2$=3.72e-05, $\hat{\text{IVAR}}$=1.31e-04, $\hat{\text{IMSE}}$=1.68e-04.}
	\label{RMDRLO}
\end{figure}

Because the likelihood is available in this example, we can implement the RM-DRLO method proposed in supplement section \ref{sec:rmdrlo}, in which an RM-DR estimate is used as a starting point to a local BFGS optimizer targeting the log likelihood. We compare the results to the MLE, which is found via a global optimization algorithm. The performance metrics for RM-DRLO shown in Fig. \ref{RMDRLO} across the parameter space are very similar to those of MLE, shown in Fig. \ref{FN_metrics}. The RM-DRLO estimator appears less variable overall, and has better estimation in some regions, such as the top left boundary. Additionally, RM-DRLO has better overall performance than MLE in terms of integrated metrics. This example illustrates that the combined estimation approach achieves comparable performance to MLE but at a lower computational cost. With a pre-learnt reconstruction map to provide good starting points for a local algorithm, RM-DRLO is more computationally efficient, with a speed over 150 times faster compared to a global approach in this case. Thus, RM-DRLO can potentially serve as a way to speed up estimation for optimization-based approaches. 

\bibliography{mref}